\documentclass[letterpaper,journal,final]{IEEEtran}

\newcommand{\subparagraph}{}

\usepackage{graphicx,psfrag}
\usepackage{amsmath,amssymb,amsfonts,mathrsfs,mathtools}
\usepackage{color,epstopdf}
\usepackage{algorithmic}
\usepackage{enumerate} 
\usepackage{subfigure}
\usepackage{floatrow}
\usepackage{float}
\usepackage{stfloats}
\usepackage{flushend}
\usepackage{bbold}
\usepackage{dsfont}
\usepackage{mathrsfs}
\usepackage{relsize}
\usepackage{relsize}
\usepackage{comment}

\newtheorem{example}{Example}
\newtheorem{rem}{Remark}
\newtheorem{assumption}{Assumption}
\newtheorem{definition}{Definition}
\newtheorem{theorem}{Theorem}
\newtheorem{lem}{Lemma}
\newtheorem{corollary}{Corollary}

\newcommand{\R}{\mathbb{R}}

\newcommand{\bbm}{\begin{bmatrix}}
\newcommand{\ebm}{\end{bmatrix}}

\def\qedp{\hspace*{\fill}~{\tiny $\blacksquare$}}
\def\be{\begin{equation}}
\def\ee{\end{equation}}
\def\ba{\begin{array}}
\def\ea{\end{array}}
\def\eqa{\begin{eqnarray}}
\def\eqe{\end{eqnarray}}

\definecolor{darkgreen}{rgb}{0.0, 0.55, 0.0}
\definecolor{amaranth}{rgb}{0.9, 0.17, 0.31}
\usepackage[prependcaption,colorinlistoftodos]{todonotes}

\def\qedp{\hspace*{\fill}~{\tiny $\blacksquare$}}
\def\qedd{\hspace*{\fill}~{\tiny $\Box$}}

\begin{document}

\title{
Enforcing contraction via data
}

\author{Zhongjie Hu, Claudio De Persis, Pietro Tesi
\thanks{Zhongjie Hu and Claudio De Persis are with
the Engineering and Technology Institute, University of Groningen, 9747AG, The Netherlands (e-mail: zhongjie.hu@rug.nl, c.de.persis@rug.nl). Pietro Tesi is with DINFO, University of Florence, 50139 Florence,
Italy (e-mail: pietro.tesi@unifi.it). *This publication is part
of the project Digital Twin with project number P18-03 of
the research programme TTW Perspective which is (partly)
financed by the Dutch Research Council (NWO). The first author is supported by the China Scholarship Council. The last author is supported by the European Union under the Italian National Recovery and Resilience Plan (NRRP) of NextGenerationEU, partnership on “Telecommunications of the Future” (PE00000001 - program “RESTART”). 
}%
}

\maketitle
\begin{abstract}
We present data-based conditions for enforcing contractivity via feedback control and obtain desired asymptotic properties of the closed-loop system. We focus on unknown nonlinear control systems whose vector fields are expressible via a dictionary of functions and derive data-dependent semidefinite programs whose solution returns the controller that guarantees contractivity. 
When data are perturbed by disturbances that are linear {\color{black}combinations} of sinusoids of known frequencies (but unknown amplitude and phase) and constants, we remarkably obtain conditions for contractivity that do not depend on the magnitude of the disturbances, with imaginable positive consequences for the synthesis of the controller. Finally,  we show how to design from data an integral controller for nonlinear systems that achieves constant reference tracking and constant disturbance rejection.
\end{abstract}

\section{Introduction}\label{sec:intro}

\IEEEPARstart{T}h{\color{black}{ERE}} is widespread attention in the control community towards new ways to utilize data collected from dynamical systems in the process of designing feedback control. 
The interest is motivated by the shared belief   that machine learning is revolutionising 
many branches of science and engineering, including automatic control, and learning to design control from data is a first step towards a thorough cross-fertilization of machine learning and control theory. 

Albeit the use of data  has always represented the backbone of control design in areas such as adaptive control, recently a new wave of solutions have focused on the so-called direct data-driven control design, which produces control policies from data collected in offline experiments without {\em explicitly} attempting to identify the system \cite{CLS02,tanaskovic2017data}. Among the various proposed solutions, the one that  relates control design to data-dependent convex programs is one of the approaches that has caught on the most in recent times. Such approach has mainly focused on linear control systems \cite{deepc, de2019formulas,berberich2020robust,van2020noisy}. 

In this paper, we are interested {\color{black}in} data-driven control for nonlinear systems. Designing feedback control from data is inevitably harder due to the varied complexity of nonlinear dynamical systems. Many of the most successful model-based nonlinear control design techniques have exploited special structures of nonlinear systems \cite{isidori1995nonlinear}, which are difficult to work with when the systems to control are hardly known. 
To date, contributions to data-driven  control design for nonlinear systems are sparse. 
In  \cite{de2019formulas}, by taking into account the impact  of the remainder as perturbations corrupting the data, a feedback controller that stabilizes an equilibrium of nonlinear systems by the  first-order approximation was designed from data. As for linear systems, the design was based on the solution of semidefinite programs and prompted the observation that this approach might be fruitful for those classes of nonlinear systems whose control design relies on convex programs. These include bilinear systems \cite{bisoffi2020data,yuan2021data} 
and polynomial systems for which \cite{dai2020semi} and \cite{guo2021data} independently proposed distinct approaches. Later, \cite{strasser2021data} remarked that the results of \cite{guo2021data} could be extended to rational systems. By Taylor's series expansion, the results on data-driven control for linear and polynomials systems can be used for general nonlinear systems by a careful analysis of the remainder \cite{guo2022data,martin2023gaussian,cheah2022robust}. 
Another way to deal with general nonlinear systems is to express the system's vector fields via dictionaries of nonlinear functions and then design controllers that {\color{black}make} the closed-loop system dominantly linear \cite{dprt2023cancellation}. When coupled with  {\color{black}coordinate} transformations, these techniques do mimic a data-dependent version of nonlinear feedback linearization \cite[Section VII.B]{dprt2023cancellation}. Data-dependent nonlinear feedback linearization has been also investigated {\color{black}in} 
\cite{alsalti2023data}, and using Gaussian process regression \cite{umlauft2019feedback}. 
Other approaches consist of looking at the nonlinear system as a linear one with state dependent matrices and recursively designing the feedback controller at each time step via linear control techniques \cite{dai2021statedependent} or 
carrying out an LPV embedding of an incremental representation of the nonlinear system followed by a data-driven control synthesis \cite{verhoek2023direct}. 

{\color{black} Our data-driven control design will be guided by the concept of contraction. Contraction has emerged as a powerful property for studying the asymptotic convergence of system trajectories with respect to one another, and has found application in many problems of estimation and control, such as tracking, synchronization, robustness and observer design \cite{lohmiller1998contraction,angeli2002lyapunov,pavlov2004convergent,sontag2010contractive}. These contraction techniques are highly relevant also for networked control, where ensuring exponential convergence and robustness across distributed agents and communication links is critical. Many related concepts have been proposed and studied \cite{ruffer2013convergent,tran2018convergence}, and it is hard to make an exhaustive survey of t}he contributions, for which we refer the reader to \cite{FB-CTDS}.
Being a powerful concept to deal with nonlinear systems analysis and control, contraction theory has been {\color{black}also} employed in learning-based control. Some existing results have been surveyed in \cite{tsukamoto2021contraction}, which uses estimation techniques based on neural networks, 
while  \cite[Section 4]{kawano2023lmi} uses Gaussian  process regression to estimate a closed-loop system and  give a condition  for incremental exponential stability in the second moment.

{\it Contribution and related works.} 
For autonomous discrete-time dynamical systems whose vector field is a contraction mapping, convergence to a fixed point is a consequence of the contraction mapping theorem. 
More in general and for both discrete- and continuous-time systems, concepts such as contraction analysis \cite{lohmiller1998contraction}, incremental stability \cite{angeli2002lyapunov} and convergent systems \cite{pavlov2004convergent} (whose relations have been studied in \cite{ruffer2013convergent}, \cite{tran2018convergence})  have established deep connections between asymptotic properties of  trajectories with respect to each other and Lyapunov analysis. When specialized to systems that have an  equilibrium, establishing contractivity does not require the explicit knowledge of the equilibrium itself, a feature which is useful when dealing with unknown systems. More importantly, contractivity is  instrumental in solving various problems such as observer design and  output regulation. 

The purpose of this paper is to present data-based conditions for enforcing contractivity via feedback control and obtain desired asymptotic properties of the closed-loop system. {\color{black}
With the exception of the papers mentioned above, to the best of our knowledge there is no result in the data-driven control design literature that deals with the problem of making a closed-loop system  contractive by feedback. The solution proposed in this paper reduces the design to solving data-based convex programs, and hence it retains the simplicity of the method introduced in \cite{de2019formulas}.  
}

As in \cite{dprt2023cancellation}, we focus on unknown nonlinear control systems whose vector fields are expressible via a dictionary of functions and aim at deriving data-dependent semidefinite programs whose solution returns the controller that guarantees contractivity {\color{black}(later we relax the complete dictionary assumption by considering remainders in the dynamics).} {\color{black}On the other hand, the design techniques presented here are noticeably different from those in \cite{dprt2023cancellation}, as we thoroughly discuss in Section \ref{subsec:comparison}. They allow us to establish contractivity of the closed-loop system and design nonlinear integral controllers for disturbance rejection and reference tracking. 
The current paper  is also distinctly new compared to other papers  by the authors and  devoted to data-driven nonlinear control design. The paper \cite{guo2021data} deals with {\em polynomial} systems and provides Sum-of-Squares (SOS) based design conditions, which are more computationally expensive than the Semidefinite Programs (SDP) used here. SOS conditions are also used to deal with general nonlinear systems in {\color{black} \cite{guo2022data} (see also \cite{guo2024data})}, {\color{black}resorting}  to a Taylor's expansion of the vector field and {\color{black}treating} the remainder as a disturbance. Other works give SDP-based results but exploit some prior knowledge about the structure of the systems to be controlled: 
\cite{bisoffi2020data} investigates the stabilization of bilinear systems along with a characterization of the region of attraction, 
in  \cite{luppi2022data} the focus is on nonlinear systems of Lur'e type, whereas  \cite{gdpt2023cdc} shows how to iteratively use data in a backstepping-like procedure to design stabilizers for systems that have a strict feedback form. 
}

Using semidefinite programs to guarantee contractivity via feedback has been used in the case the system's model is known  \cite{pavlov2005convergent,manchester2017control,andrieu2019lmi}, but the use of data to offset the uncertainty on the model makes the presented results distinctively novel. The presented approach also carries over to the case of data that are perturbed by unknown but bounded process disturbances. Even more, if the disturbances are linear combinations of  constant and sinusoidal  signals of known frequencies (but unknown amplitude and phase), we present a new result where the data-based conditions for contractivity do not depend on the magnitude of the disturbance  affecting the data and  remarkably bear similarities with the ones obtainable in the case of noise-free data.  {\color{black}This has} positive consequences for the synthesis of the controller, {\color{black}because the data-dependent convex program that returns the controller is less complex than in the case of data perturbed by general noise (cf.~the constraints \eqref{eq:SDP2} and \eqref{eq:SDP2-noise})}. 

As our final contribution, in order to underscore the relevance of contractivity in data-driven control problems different from mere stabilization,  we show how to design from data an integral controller for nonlinear systems that achieves constant reference tracking and constant disturbance rejection. For this problem, we follow the idea in \cite{giaccagli2020sufficient}, \cite{giaccagli2023lmi} of rendering  the given system, extended with the integral controller and without forcing inputs, contractive and then analyze the impact of the reference and the disturbance by relying on the established contractivity. This idea does not rely on special forms of the system and is therefore well suited for data-based control design.

{\it Outline.} 
The framework and problem formulation are discussed in Section \ref{sec:framework}. The main result is given in Section \ref{sec:exact}, which is then expanded to deal with noisy data (Section \ref{sec:noisy-data}). The special case of disturbances given by linear {\color{black}combinations} of sinusoids and constants is studied in Section \ref{sec:dist-known-freq}. The paper ends with the application of the previous results to the design of integral controllers (Section \ref{sec:integral-control}). To make the paper self-contained, we added an Appendix that contains a few basic results about contractive and convergent systems that are used in the paper. 

\textit{Notation.} Throughout the paper, $\mathbb{R}$ denotes the set of real numbers, while $\mathbb{R}_{> c}$ ($\mathbb{R}_{\ge c}$) denotes the set of real numbers greater than (greater than or equal to) the real number $c$. Given a symmetric matrix $M$, $M \succ 0$ ($M\succeq0$) indicates that $M$ is positive definite (positive semidefinite), while $M \prec 0$ ($M\preceq0$) indicates that $M$ is negative definite (negative semidefinite).  $\mathbb S^{n \times n}$ denotes the set of real-valued symmetric matrices. We abbreviate a symmetric matrix $\left[\begin{smallmatrix} M & N \\ N^\top & P\end{smallmatrix}\right]$ as $\left[\begin{smallmatrix} M & \star \\ N^\top & P\end{smallmatrix}\right]$. {\color{black}To shorten some formulas, we denote $M+M^\top$ as $\text{Sym}(M)$. The notation $\text{diag}(M_1, \ldots, M_\mu)$ indicates a diagonal or a block diagonal matrix depending on whether the quantities $M_1, \ldots, M_\mu$ are all scalar or not. }Finally, we let $|v|$ be the 2-norm of a vector $v$,  $\lVert M \rVert$ be the induced 2-norm of a matrix $M$, and $\lambda_{\min}(M)$ ($\lambda_{\max}(M)$) be the smallest (largest) eigenvalue of $M$. 


\section{Set-up and problem formulation} \label{sec:framework}

We consider a continuous-time system of the form
\begin{equation}
\label{system}
\dot x = f(x) + B u
\end{equation}  
where $x \in \mathbb R^{n}$ is the state, $u \in \mathbb R^{m}$ is the control input\footnote{ {\color{black}Here and elsewhere in the paper, by an abuse of notation, $x$ and $u$ may both denote functions of time as well as their evaluation at time $t$.}}. The vector field $f(x)$ and the input matrix $B$ are unknown. To ease the analysis we focus on the case of a constant input matrix $B$. 
We are interested in the problem of making the system contractive via feedback. We will introduce later a problem that motivates the interest in this property.  

\smallskip

Without further priors about the vector field $f$, any control design is challenging. Hence, we introduce the following:
\begin{assumption} \label{ass:Z}
We know a continuously differentiable vector-valued function $Z: \mathbb R^n \rightarrow \mathbb R^s$ of the form
\begin{equation} \label{eq:Z}
Z(x) = \begin{bmatrix} x \\ Q(x) \end{bmatrix}
\end{equation}
with $Q: \mathbb R^n \rightarrow \mathbb R^{s-n}$ containing only nonlinear functions, such that 
$f(x) = AZ(x)$ for some matrix $A \in \mathbb R^{n \times s}$. \qedd
\end{assumption} 

Under Assumption \ref{ass:Z}, system \eqref{system} can be  equivalently written as
\begin{equation}
\label{system2}
\dot x = A Z(x) + B u
\end{equation}  
with $A,B$ unknown. {\color{black} Later in Section \ref{sec:noisy-data} we will discuss how to relax Assumption \ref{ass:Z}.}

\noindent \emph{Dataset.} The lack of knowledge about $A,B$ are offset by data collected in an experiment, which returns the following dataset to the designer:
\begin{equation} \label{dataset}
\mathbb D := \left\{ (x_i,u_i, \dot x_i) \right\}_{i=0}^{T-1}
\end{equation} 
where $T>0$ is an integer equal to  the number of samples of the dataset, $x_i:= x(t_i), u_i:=u(t_i), \dot x_i:= \dot x(t_i)$ and {\color{black}$0\le t_0< t_1<\ldots< t_{T-1}$} are the sampling times. No other requirement is imposed on the sampling times {\color{black} other than  that  they be distinct}. By construction, the samples satisfy the relation  $\dot x_i=AZ(x_i)+Bu_i$ for $i=0,\ldots,T-1$, $T>0$. Note that,  $Z(x_i)$ can be computed from $x_i$ and {\color{black}the function $Z$}, since the latter  is known. {\color{black} The computed samples $Z(x_i)$, for $i=0,\ldots,T-1$, must possess sufficient information regarding the dynamics of the system, in a sense that we will comment on 
at the beginning of Section \ref{subsec:discussion}
below.
}

We also assume that the time derivative of the state $\dot x_i$ can be measured. {\color{black}When this is not possible}, there are two ways to relax it. One is to  consider approximate estimates of the derivatives, regard the estimation error as a perturbation on the data and design the controller accordingly (see Section \ref{sec:noisy-data} on how to establish contractivity with noisy data). The other is to use a different data collection scheme that considers integral versions of the relation \eqref{system2} (see \cite[Appendix A]{de2023event} for details). Finally, we begin to study the scenario in which data are not perturbed by disturbances or noise. 

\smallskip

\noindent \emph{Problem.}~We are interested in the design of a feedback controller of the form $u=K Z(x)$, which makes the closed-loop system 
\begin{equation} \label{eq:closed}
\dot x = (A+BK) Z(x)
\end{equation}
contractive on a set $\mathcal{X}\subseteq \mathbb{R}^n$. We recall below such a property in a form that is suitable for our purposes:
\begin{definition}
Consider a set $\mathcal{X}\subseteq \mathbb{R}^n$.  
System \eqref{eq:closed} is exponentially contractive on $\mathcal{X}$ if 
\be\label{contractivity}
\hspace{-3mm}
\begin{array}{l}
{\color{black}\exists P \in \mathbb S^{n \times n}, P \succ 0}, \beta>0 \textrm{ such that } \forall x\in  \mathcal{X}\\
((A+BK)\displaystyle\frac{\partial Z{\color{black}(x)}}{\partial x})^\top P^{-1}+P^{-1}
(A+BK)\displaystyle\frac{\partial Z{\color{black}(x)}}{\partial x}\preceq -\beta {\color{black}P}^{-1}. 
\end{array}
\ee
The set $\mathcal{X}$ is a contraction region with respect to the metric ${\color{black}P}^{-1}$. \qedd
\end{definition}

{\color{black} The terminology adopted in the definition  is consistent with those found in e.g.~\cite{duvall2024global,aylward2008stability} (see a more detailed discussion in \cite{hu2025data}, before Proposition 1).} 
{\color{black} The contractivity property is}
instrumental in several  estimation and control problems, whose use in the control literature has been popularized by \cite{lohmiller1998contraction}.  In the original definition therein, the metric ${\color{black}P}^{-1}$ is taken as state- and time-dependent. Since the search for a state- or time-dependent metric is difficult, here we concentrate our efforts on a constant metric. In fact, to complicate the situation here is the lack of knowledge of the matrices $A,B$ in \eqref{system2} and the presence of the design variable $K$. Hence, beside ${\color{black}P}$ and $\beta$, one has to simultaneously seek  a $K$ that enforces \eqref{contractivity}.  Finally, in Section \ref{sec:integral-control} we will consider a tracking  problem with constant reference and disturbance signal of arbitrary magnitude, for which considering a constant metric results in no loss of generality \cite[Theorem 2]{duvall2024global}. 

{\color{black}In the next section, we give a solution to the problem of designing a feedback control from data that makes the closed-loop system contractive. To start, the solution makes use of noise-free data to highlight the proposed design strategy, compare with other approaches and  avoid the intricacies of noisy data. 
We will then look at the problem  of enforcing contractivity using noisy data in Section \ref{sec:noisy-data}, by which we mean that, during the phase of data acquisition,  the system \eqref{system2} is affected by an  additive perturbation, as described by the equation $\dot x = A Z(x)+Bu +Ed$, where $d$ models the unmeasured perturbation and $E$ is a matrix that specifies which parts of the dynamics are affected by the perturbation. This analysis 
also provides a framework that allows one to  address the case in which $f(x)$ does {\em not} lie in the finite-dimensional space spanned by the functions in $Z(x)$, but it is expressed as $f(x)=AZ(x)+Ed(x)$, where $d(x)$ represents a remainder, thus lifting Assumption \ref{ass:Z}. Among the control problems for which contractivity is instrumental in their solution, the output regulation problem stands out for its importance. As anticipated above, {\color{black} integral controllers are} studied in Section \ref{sec:integral-control}. As customary in output regulation problems, the perturbation affecting the plant 
{\color{black}are generated} by exosystems. Hence, in Section \ref{sec:dist-known-freq} we enrich and complement the findings of  Section \ref{sec:noisy-data} showing how this class of disturbances can be effectively filtered out by a suitable modification of the convex programs used to make a system contractive. 
}


%
%

%

\section{Control design}
\label{sec:exact}

\subsection{A data-dependent representation of the closed-loop system}

A first step we take toward designing the controller is to obtain a data-dependent representation of the closed-loop system. To this end, we arrange the dataset in the following matrices:
\begin{subequations}
\label{eq:data}
\begin{align}
&U_0 := \begin{bmatrix} u_0 & u_1 & \cdots & u_{T-1}  \end{bmatrix} 
\in \mathbb R^{m \times T} \,, \label{eq:data1} \\
& X_0 := \begin{bmatrix} x_0 & x_1 & \cdots & x_{T-1}  \end{bmatrix} 
\in \mathbb R^{n \times T} \,, \label{eq:data2} \\
& X_1 := \begin{bmatrix} \dot x_0 & \dot x_1 & \cdots & \dot x_{T-1}  \end{bmatrix} 
\in \mathbb R^{n \times T} \,, \label{eq:data3} \\
& 
Z_0 := \begin{bmatrix} x_0 & x_1 & \cdots & x_{T-1}   \\ 
Q(x_0) & Q(x_1) & \cdots & Q(x_{T-1}) 
\end{bmatrix}  \in \mathbb R^{s \times T}.   \label{eq:data4}
\end{align}
\end{subequations}
Note that these matrices satisfy the identity
\[
X_1 =A Z_0 + B U_0.  
\]
The following result, established in \cite[Lemma 1]{dprt2023cancellation} and recalled here without proof, returns the data-dependent representation of the closed-loop system we are interested in. 
\begin{lem} \label{lem:main}
Consider any matrices $K \in \mathbb R^{m \times s}$,
$G \in \mathbb R^{T \times s}$ such that
\begin{equation} \label{eq:GK}
\begin{bmatrix} K \\ I_s \end{bmatrix} = \begin{bmatrix} U_0 \\ Z_0 \end{bmatrix} G \,.
\end{equation}
Let $G$ be partitioned as $G = \begin{bmatrix} G_1 & G_2 \end{bmatrix}$, where
$G_1 \in \mathbb R^{T \times n}$ and $G_2 \in \mathbb R^{T \times (s-n)}$.
Let Assumption \ref{ass:Z} hold. Then system \eqref{system} under the control law $u=KZ(x)$ results in
the closed-loop dynamics 
\begin{equation} \label{eq:GK_closed}
\dot x = (A+BK) Z(x)=Mx + N Q(x) 
\end{equation}
where $M:=X_1 G_1$ and $N:=X_1 G_2$.
\end{lem} 

\emph{Proof.} See \cite[Lemma 1]{dprt2023cancellation}. 
\qedp

\smallskip

We note that the right-hand side of {\color{black}the condition \eqref{eq:GK} and the equation \eqref{eq:GK_closed}} only depend on the matrices of data $U_0, Z_0, X_1$ and the decision variable $G$. Hence, we will design $G$ that enforces the desired 
contractivity 
property and then obtain the control gain $K$ via \eqref{eq:GK}. 


\subsection{Contractivity from data} 
%
The first statement {\color{black}below} gives data-dependent conditions under which there {\color{black}exist} ${\color{black}P}$, $K$ and {\color{black}$\beta$} for which \eqref{contractivity} holds. {\color{black}The statement, which is the main result of this section, shows that, given a Lipschitz bound on the nonlinear term $Q(x)$ on the set $\mathcal{X}$, the data-based design of a nonlinear feedback control that makes the closed-loop system contractive is reduced to an LMI, hence to a feasibility problem of a convex program. Afterwards, we give a few remarks that discuss some technical  aspects of the result, 
argue that the LMI and its feasibility are related to  robust control design problems,
draw  consequences on the asymptotic properties of the closed-loop system, and compare the result with other data-driven nonlinear control design techniques. }

\begin{theorem} \label{thm:contractivity}
Consider the nonlinear system  \eqref{system}. Let Assumption \ref{ass:Z} hold and $Z(x)$ be of the form \eqref{eq:Z}. 
Let 
$ R_Q\in \mathbb R^{n \times {\color{black}\tau}}$ be a known matrix and $\mathcal{X}\subseteq \mathbb{R}^n$ a set such that
\be\label{asspt}
\frac{\partial Q}{\partial x}(x)^\top \frac{\partial Q}{\partial x}(x)\preceq R_Q R_Q^\top \text{ for any $x\in \mathcal{X}$}.
\ee
Consider 
the following {\color{black}SDP} in the decision variables ${\color{black}P} \in \mathbb S^{n \times n}$,
$Y_1 \in \mathbb R^{T \times n}$, 
$G_2 \in \mathbb R^{T \times (s-n)}$ and $\alpha \in \mathbb R_{> 0}$:
\begin{subequations}
\label{eq:SDP}
\begin{alignat}{6}
& {\color{black}P} \succ 0, \label{eq:SDP0}\\
& Z_0 Y_1 = \begin{bmatrix} {\color{black}P} \\ 0_{(s-n) \times n} \end{bmatrix} \,,
\label{eq:SDP1} \\
& \begin{bmatrix} X_1 Y_1 +(X_1 Y_1)^\top + \alpha I_n  & 
{\color{black}\star} 
& 
{\color{black}\star} 
\\ 
{\color{black}G_2^\top X_1^\top}  & - I_{s-n} & 
{\color{black}\star} 
\\
{\color{black} R_Q^\top P} & 0 & - I_{{\color{black}\tau}}
\end{bmatrix}  \preceq 0 \,, \label{eq:SDP2} \\
& Z_0 G_2 = \begin{bmatrix} 0_{n \times (s-n)} \\ I_{s-n} \end{bmatrix} \,.
\label{eq:SDP4}
\end{alignat}
\end{subequations} 
If the program is feasible then the control law 
\be\label{state-feedback}
u=KZ(x)
\ee 
with 
\begin{eqnarray} \label{eq:K_SDP}
K= U_0 \begin{bmatrix} Y_1 {\color{black}P}^{-1}
& G_2 \end{bmatrix}
\end{eqnarray}
is such that  the closed-loop dynamics \eqref{system}, \eqref{state-feedback} {\color{black} are} exponentially contractive on $\mathcal{X}$, i.e. \eqref{contractivity} holds. 
\end{theorem} 


\emph{Proof.} If  \eqref{eq:SDP} is feasible, then we can define  $G_1:=Y_1{\color{black}P}^{-1}$
and note that \eqref{eq:SDP1}, \eqref{eq:SDP4}  and  the definition of $K$ in  \eqref{eq:K_SDP},  give \eqref{eq:GK}. Hence, by Lemma \ref{lem:main},  the closed-loop dynamics $\dot x = (A+BK) Z(x)$ can be written as $\dot x = Mx + N Q(x)$, with 
$M=X_1G_1$ and $N=X_1G_2$. 
By Schur complement, \eqref{eq:SDP2}  gives 
\[
\begin{bmatrix} X_1 Y_1 +(X_1 Y_1)^\top + {\color{black}P} R_Q R_Q^\top {\color{black}P}+ \alpha I_n  &{\color{black}\star}  \\ 
{\color{black}G_2^\top X_1^\top}  & -I_{s-n} 
\end{bmatrix}  \preceq 0 \,. \label{} 
\]
Another Schur complement returns 
\[
X_1 Y_1 +(X_1 Y_1)^\top + \alpha I_n +  {\color{black}P} R_Q R_Q^\top {\color{black}P}+ X_1 G_2 (X_1 G_2)^\top  \preceq 0 \,. \label{} 
\]
Left- and right-multiply by ${\color{black}P}^{-1}$ and recall that $G_1 := Y_1 {\color{black}P}^{-1}$. Then 
\be\label{int-ineq}
\ba{l}
{\color{black}P}^{-1} X_1 G_1 +(X_1 G_1)^\top {\color{black}P}^{-1}  + \alpha {\color{black}P}^{-2}  +  R_Q R_Q^\top \\[1mm]
 \hspace{3.5cm} + {\color{black}P}^{-1} X_1 G_2 (X_1 G_2)^\top {\color{black}P}^{-1}  \preceq 0 \,. 
\ea
\ee
{\color{black}
This shows that there exist a real number $\lambda =1$ and  matrices 
\[\ba{rl}
H :=& {\color{black}P}^{-1} X_1 G_1 +(X_1 G_1)^\top {\color{black}P}^{-1}  + \alpha {\color{black}P}^{-2}\\ 
J :=& {\color{black}P}^{-1} X_1 G_2\\ 
L :=& I_n\\
O :=& R_Q R_Q^\top
\ea\] 
that  satisfy $H+\lambda JJ^\top +\lambda^{-1} L^\top O L\preceq 0$. 
If the solution of \eqref{eq:SDP} is such that $J=0$, then 
$(A+BK)\frac{\partial Z}{\partial x}= X_1 G_1$ and \eqref{int-ineq} implies \eqref{contractivity} with $\beta:=\alpha \lambda_{\min}({\color{black}P}^{-1})$, where $\lambda_{\min}({\color{black}P}^{-1})$ is the minimum eigenvalue of ${\color{black}P}^{-1}$, which ends the proof.  
On the other hand, if the solution of \eqref{eq:SDP} is such that $J\ne 0$, 
then by the nonstrict Petersen's lemma \cite[Fact 2]{bisoffi2022data}, we  have that $H+\lambda JJ^\top +\lambda^{-1} L^\top O L\preceq 0$ implies $H+ J R^\top  L+
L^\top R J^\top\preceq 0$ for all $R\in \mathcal{R}:=\{R\colon R R^\top\preceq O\}$, i.e. 
\[\ba{l}
{\color{black}P}^{-1} X_1 G_1 +(X_1 G_1)^\top {\color{black}P}^{-1}  + \alpha {\color{black}P}^{-2}
\\
\hspace{3cm}
+{\color{black}P}^{-1} X_1 G_2 R^\top + R G_2^\top X_1^\top  {\color{black}P}^{-1}\preceq 0,\\[2mm]
\text{ for all }
R\in \mathcal{R}:=\{R\colon RR^\top\preceq R_Q R_Q^\top\}.
\ea\]
}
By \eqref{asspt}, for any $x\in \mathcal{X}$, the inequality above gives
\[
\ba{l}
{\color{black}P^{-1} X_1 G_1 +(X_1 G_1)^\top P^{-1}  + \alpha P^{-2}  +  \frac{\partial Q}{\partial x}(x)^\top \frac{\partial Q}{\partial x}(x)} \\[1mm]
 \hspace{3.5cm}{\color{black} + P^{-1} X_1 G_2 (X_1 G_2)^\top P^{-1}  \preceq 0}
 \ea
\]
{\color{black}and further}
\be\label{eq:conseq-main-cond-for-contrcvty}\ba{l}
{\color{black}\text{Sym}(P^{-1} X_1 G_1 + P^{-1}X_1 G_2 \displaystyle \frac{\partial Q}{\partial x}(x)) + \alpha P^{-2}\preceq 0, } 
\ea\ee
that is, recalling that $\beta=\alpha \lambda_{\min}({\color{black}P}^{-1})$, 
\[
\ba{l}
{\color{black}P}^{-1} X_1 G \frac{\partial Z(x)}{\partial x}+
\frac{\partial Z(x)}{\partial x}^\top G^\top X_1^\top {\color{black}P}^{-1}\preceq -\beta {\color{black}P}^{-1} \\ \hspace{5cm}  \text{ for all $x\in \mathcal{X}$},
\ea
\]
which ends the proof. \qedp

\subsection{{\color{black}Discussion}} \label{subsec:discussion}
In this subsection, we give a few remarks that discuss some technical aspects of Theorem \ref{thm:contractivity}, 
argue that the LMI and its feasibility are related to  robust control design problems,
and draw consequences on the asymptotic properties of the closed-loop system.

{\color{black} 
\subsubsection*{On the persistence of excitation}  The data-dependent representation of the closed-loop system, based on which its contractivity is enforced,  is obtained by imposing the consistency relations \eqref{eq:SDP1}, \eqref{eq:SDP4} and designing the controller gain $K$ to lie in the span of $U_0$ by \eqref{eq:K_SDP}. In this way, the condition \eqref{eq:GK}  is satisfied and Lemma \ref{lem:main} holds. Note that the consistency relations \eqref{eq:SDP1}, \eqref{eq:SDP4} require $Z_0$ to have full row rank. Apart from a few exceptions, such as \cite[Theorem 5]{alsalti2023design}, the problem of designing an input sequence $u_0, u_1, \ldots, u_{T-1}$ such that the matrix $Z_0$ has full row rank has not been fully understood yet. This is in contrast with the case of linear systems, where the use of a persistently exciting input sequence 
during the data acquisition phase guarantees that the matrix $X_0$ (the counterpart of $Z_0$ in the context of linear systems)  has such a property \cite{willems2005note}, and this was used in the data-based design of linear control systems in \cite{de2019formulas}.  
}

\subsubsection*{On condition \eqref{asspt}}
Theorem \ref{thm:contractivity} holds under the local Lipschitz property \eqref{asspt}. The result also holds for other types of  nonlinearities as shown below.
\begin{enumerate}
\item[(1)] \emph{More general nonlinearities.} 
These are nonlinearities that satisfy more general conditions such as 
\be\label{asspt-general}
\ba{l}
\displaystyle\frac{\partial Q}{\partial x}(x)^\top R \frac{\partial Q}{\partial x}(x)+S \frac{\partial Q}{\partial x}(x)\\[2mm]
\hspace{2.5cm}
\displaystyle+\frac{\partial Q}{\partial x}(x)^\top S^\top\preceq W \text{ for any $x\in \mathcal{X}$}
\ea
\ee
where $W=W^\top \succeq 0$,  $R=R^\top\succeq 0$, $S$  are known matrices, provided that the LMI \eqref{eq:SDP2}  is replaced by 
\[
\begin{bmatrix} X_1 Y_1 +(X_1 Y_1)^\top + \alpha I_n  & 
{\color{black}\star}
& 
{\color{black}\star}
\\ 
{\color{black}G_2^\top X_1^\top - S^\top P} & - R & 
{\color{black}\star} 
\\
{\color{black} W^{1/2}P} & 0 & - I_{\tau}
\end{bmatrix}  \preceq 0.
\]
{\color{black}
In fact, the Schur complement applied to the inequality above returns
\[
\begin{bmatrix} X_1 Y_1 +(X_1 Y_1)^\top + {\color{black}P} W {\color{black}P}+ \alpha I_n  & {\color{black}\star}
\\ 
{\color{black}G_2^\top X_1^\top - S^\top P}  & -R
\end{bmatrix}  \preceq 0 \,. \label{} 
\]
By left- and right-multiplying by 
${\color{black}{\rm diag}}({\color{black}P}^{-1}, I_{s-n})$ and setting $G_1 := Y_1 {\color{black}P}^{-1}$, the inequality above is rewritten as 
\[
\!\!
\begin{bmatrix} {\color{black}P}^{-1} X_1 G_1 +({\color{black}P}^{-1} X_1 G_1)^\top + W + \alpha {\color{black}P}^{-2}  & \!\!\!\!
{\color{black}\star}
\\ 
{\color{black}G_2^\top X_1^\top P^{-1} - S^\top}  & -R
\end{bmatrix}  \preceq 0. \label{} 
\]
Left-multiplying the last inequality by $\left[\begin{smallmatrix} I  \\ \frac{\partial Q}{\partial x}\end{smallmatrix}\right]^\top$, 
right-multiplying it by $\left[\begin{smallmatrix} I  \\ \frac{\partial Q}{\partial x}\end{smallmatrix}\right]$,
and bearing in mind \eqref{asspt-general}, it follows that 
\[
\left[\begin{matrix} I  \\ \frac{\partial Q}{\partial x}\end{matrix}\right]^\top\!\!
\begin{bmatrix} {\color{black}P}^{-1} X_1 G_1 +({\color{black}P}^{-1} X_1 G_1)^\top + \alpha {\color{black}P}^{-2}  & 
\star
\\ 
{\color{black}G_2^\top X_1^\top P^{-1}}  & 0
\end{bmatrix}\!\!\left[\begin{matrix} I  \\ \frac{\partial Q}{\partial x}\end{matrix}\right]\!\!\preceq \!\!
0\]
for all $x\in \mathcal{X}$, which is \eqref{contractivity}, as claimed. }
Examples of nonlinearities that satisfy \eqref{asspt-general} include those that satisfy incremental sector bound conditions \cite[Lemma 4.4]{dalto2013incremental}. 

{\color{black} \item[(2)] \emph{Monotonic nonlinearities.} If the nonlinearities in $Q(x)$ are of monotonic type \cite{giaccagli2023lmi}, Theorem \ref{thm:contractivity} yields a controller that assigns an exponentially contractive dynamics to the closed-loop system  {\it without} requiring the linear part to dominate the growth of $Q(x)$ but rather exploiting the monotonic property of $Q(x)$. Recall that monotonic nonlinearities satisfy the inequality \eqref{asspt-general} with $R=0$ and $W=0$, that is, they satisfy \be\label{asspt-general-monotonic}
S \frac{\partial Q}{\partial x}(x)
\displaystyle+\frac{\partial Q}{\partial x}(x)^\top S^\top\preceq 0. 
\ee
In this case, 
Theorem \ref{thm:contractivity} holds under the SDP \eqref{eq:SDP}, provided that the condition \eqref{eq:SDP2} is replaced by 
\[
X_1 Y_1 +  Y_1^\top X_1^\top \preceq -\alpha I_n \text{ and } X_1 G_2 =P S.
\]
{\color{black} In fact, the above condition is equivalent to
\[
\begin{bmatrix} P^{-1} X_1 G_1 +(P^{-1} X_1 G_1)^\top + \alpha P^{-2}  & \star \\ 
G_2^\top X_1^\top P^{-1} -S^\top  & 0
\end{bmatrix}\preceq 0.
\]
Left-multiplying the last inequality by $\left[\begin{smallmatrix} I  \\ \frac{\partial Q}{\partial x}\end{smallmatrix}\right]^\top$, 
right-multiplying it by $\left[\begin{smallmatrix} I  \\ \frac{\partial Q}{\partial x}\end{smallmatrix}\right]$,
and bearing in mind 
\[
\left[\begin{matrix} I  \\ \frac{\partial Q}{\partial x}\end{matrix}\right]^\top\!\!
\begin{bmatrix} 0  & S
\\ 
S^\top  & 0
\end{bmatrix}\!\!\left[\begin{matrix} I  \\ \frac{\partial Q}{\partial x}\end{matrix}\right] \preceq 
0,
\]
it follows that 
\[
\left[\begin{matrix} I  \\ \frac{\partial Q}{\partial x}\end{matrix}\right]^\top\!\!
\begin{bmatrix} P^{-1} X_1 G_1 +(P^{-1} X_1 G_1)^\top + \alpha P^{-2}  & \star  \\ 
G_2^\top X_1^\top P^{-1}& 0
\end{bmatrix}\!\!\left[\begin{matrix} I  \\ \frac{\partial Q}{\partial x}\end{matrix}\right]\preceq 
0
\]
for all $x\in \mathcal{X}$, which is \eqref{contractivity}, as claimed.}
The resulting controller $u=KZ(x)$, with $K$ as in \eqref{eq:K_SDP}, assigns to the nonlinear part of the closed-loop system a value ($P S$)  that contributes a nonpositive quantity to the left-hand side of the inequality 
\eqref{eq:conseq-main-cond-for-contrcvty}.

\item[(3)] \emph{$\frac{\partial Q}{\partial x}(x)$ in a convex hull.} The condition \eqref{asspt} on $\frac{\partial Q}{\partial x}(x)$ is a ``global" condition that must hold {\color{black}at any point of }
$\mathcal{X}$. Replacing this global bound with local ones might give computational advantages. This can be achieved if there exist matrices $Q_i$, $i=1,2,\ldots, \nu$, and functions $\lambda_i(x)$,  $i=1,2,\ldots, \nu$, such that \cite{pavlov2005convergent,barabanov2019contraction}
\[\ba{l}
\displaystyle\frac{\partial Q}{\partial x}(x) = \sum_i^\nu \lambda_i(x) Q_i, \; \lambda_i(x)\ge 0,\\ 
\hspace{2cm}\sum_i^\nu \lambda_i(x) =1, \text{ for all $x\in \mathcal{X}$.} 
\ea\]
Then Theorem \ref{thm:contractivity} can be reformulated through 
the $\nu$ ``local" conditions 
\[
\begin{bmatrix}
X_1 Y_1 + (X_1 Y_1)^\top +\beta P & \star & \star \\
{\color{black}G_2^\top X_1^\top} & - I_{s-n} & \star \\
Q_i P & 0_{(s-n) \times (s-n)} & - I_{s-n}
\end{bmatrix}\preceq 0,
\]
instead of the single global one \eqref{eq:SDP2}. 
 Applying the Schur complement to the inequality above returns
\[
\ba{l}
X_1 G_1 P^{-1} +P^{-1} (X_1 G_1)^\top +\beta P^{-1} \\ \hspace{2cm}
+(X_1 G Q_i)^\top P^{-1}+P^{-1}
X_1 G_2 Q_i \preceq 0
\ea
\]
for  $i=1,\ldots, \nu$. Then, it follows that
\[
\ba{l}
\displaystyle\sum_{i=1}^\nu \lambda_i(x) (X_1 G_1 P^{-1} +P^{-1} (X_1 G_1)^\top +\beta P^{-1} \\ \hspace{2cm} +(X_1 G Q_i)^\top P^{-1}+P^{-1}
X_1 G_2 Q_i) \preceq 0
\ea
\]
for all $x\in \mathcal{X}$, and bearing in mind the expression of $\displaystyle\frac{\partial Q}{\partial x}(x)$, we arrive at \eqref{contractivity}, as claimed.
}
\end{enumerate}

\smallskip  

\subsubsection*{On condition \eqref{eq:SDP}} Contractivity and the asymptotic properties given above are guaranteed by \eqref{eq:SDP}. {\color{black}We have previously remarked that} the conditions \eqref{eq:SDP1}, \eqref{eq:SDP4} are used to obtain the data-dependent representation \eqref{eq:GK_closed} of the closed-loop system: $\dot x = Mx + N Q(x)$,
where $M=X_1 G_1$ and $N=X_1 G_2$. On the other hand, to enforce contractivity on $\mathcal{X}$, i.e. to enforce the inequality (cf.~{\color{black}\eqref{eq:conseq-main-cond-for-contrcvty}})
\[\ba{l}
{\color{black}\text{Sym}(P^{-1} X_1 G_1 + {\color{black}P}^{-1}X_1 G_2 \displaystyle \frac{\partial Q}{\partial x}(x)) + \alpha {\color{black}P}^{-2}\preceq 0}
\ea
\]
\!\!the key property 
is \eqref{eq:SDP2}, 
which, written as
\[\ba{l}
{\color{black}P}^{-1} X_1 G_1 +(X_1 G_1)^\top {\color{black}P}^{-1}  + \alpha {\color{black}P}^{-2}  +  R_Q R_Q^\top \\
\hspace{3.5cm}+ {\color{black}P}^{-1} X_1 G_2 (X_1 G_2)^\top {\color{black}P}^{-1}  \preceq 0 \,, 
\ea\]
reveals the aim of our control design. In fact, the possibility of fulfilling this condition is  related to the existence of a matrix $ P \succ 0$ and 
for the linear part of the controller ${\color{black}P}^{-1} X_1 G_1 +(X_1 G_1)^\top {\color{black}P}^{-1}$ to dominate both ${\color{black}P}^{-1} X_1 G_2 (X_1 G_2)^\top {\color{black}P}^{-1}$ and the contribution of the Jacobian 
$\frac{\partial Q}{\partial x}(x)$, 
which appears through the term 
 ${R}_Q {R}_Q^\top$. Controllability of the pair $(\overline A,B)$, where $A=\begin{bmatrix}\overline A & \hat A\end{bmatrix}$, $\overline A\in \mathbb{R}^{n\times n}$, also plays a role in achieving this goal. 

\subsubsection*{Feasibility of \eqref{eq:SDP}} We discuss now the feasibility of the SDP. The first observation is that the best one can hope for is that the problem of ensuring contractivity is feasible when we have {\em complete information} on the system dynamics, i.e. in the model-based case. Let $Z_m$ be ``the'' minimal vector-valued function that describes the dynamics of the system, namely $f(x)=A_mZ_m(x)$ and no other vector-valued function $Z$ of dimension smaller than $Z_m$ exists such that $f(x)=A Z(x)$. (Here, we use the subscript ``\emph{m}'' in $Z_m$ and $A_m$ to highlight that these quantities are model-based.)
In this case, the problem of interest is that of finding a tuple $(P_m,K_m,\beta)$, with {\color{black}$P_m \in \mathbb S^{n \times n}, P_m \succ 0$} and $\beta > 0$, such that
\begin{equation} \label{eq2}
\ba{l}
P_m^{-1} (A+BK_m) \displaystyle\frac{\partial Z_m(x)}{\partial x} + \displaystyle\frac{\partial Z_m(x)}{\partial x}^\top (A+BK_m)^\top P_m^{-1} 
\\[2mm]
\hspace{2.5cm} \preceq -\beta P_m^{-1}
\;\text{for all } x \in \mathcal X. 
\ea
\end{equation}
Suppose that $Z_m(x) =[\,x^\top \; Q_m(x)^\top\,]^\top$, where $Q_m$ collects all the nonlinearities. Partition $A_m = [A_{m1} \,\, A_{m2}]$, 
with $A_{m1}\in \R^{n\times n}$,  and
$K = [K_{m1} \,\, K_{m2}]$.
Similar to Theorem \ref{thm:contractivity},  a sufficient condition
for \eqref{eq2} to hold is that there exists a tuple $(P_m,K_{m1}, K_{m2}, \beta)$, 
with {\color{black}$P_m \in \mathbb S^{n \times n}, P_m \succ 0$} and $\beta > 0$, such that
\begin{equation} \label{eq3}
\Phi_1 P_m + P_m \Phi_1^\top + 
{\color{black} \beta}
I_n +  P_m R_m R_m^\top P_m + \Phi_2 \Phi_2^\top \preceq 0
\end{equation}
where $\Phi_k :=A_{mk}+B K_{mk}$, $k=1,2$, and $R_m$ is a bound on $\frac{\partial Q_m}{\partial x}(x)$ in the sense that 
\begin{equation} \label{eq4}
\frac{\partial Q_m}{\partial x} (x)^\top \frac{\partial Q_m}{\partial x} (x) \preceq R_m R_m^\top \quad \text{for all } x \in \mathcal X. 
\end{equation} 
Seeking a tuple $(P_m,K_{m1}, K_{m2}, \beta)$ such that 
\eqref{eq3} holds can be turned into a convex program. We say that the problem of making a closed-loop system contractive in case of complete information is feasible if a solution to \eqref{eq3} exists. We shall assume that the problem is feasible in the case of complete information, otherwise it makes no sense to seek a data-driven solution.  
We then denote by $\mathcal{K}_m$ the set of all controllers that guarantee contractivity in the sense of \eqref{eq3}, namely
\begin{equation} \label{eq5}
\ba{l}
\!\!\!\!\mathcal{K}_m := \{ K_m: \eqref{eq3} \text{ holds for some } 
P_m \in \mathbb S^{n\times n}, P_m\succ 0, \\ \hspace{6cm}
\beta > 0 \}.
\ea
\end{equation}
Now we turn our attention to the case of incomplete information, that is, when the open-loop dynamics is given by $AZ(x)+Bu$, where $A,B$ are unknown and the number of functions in $Z(x)$ is not necessarily ``minimal" ($Z(x)$ includes the functions in $Z_m(x)$ and possibly others).  Without loss of generality, assume $Z(x) =[\,x^\top \; Q_m(x)^\top \; Q_e(x)^\top\,]^\top$ where $Q_e: \mathbb R^n \rightarrow \mathbb R^e$ are the extra functions included in $Z(x)$. Hence, in general the bound $R_Q$ on $\frac{\partial Q}{\partial x}(x)$ in \eqref{asspt}, where $Q(x) =[Q_m(x)^\top \; Q_e(x)^\top\,]^\top$, will be different from the bound 
$R_m$ on $\frac{\partial Q_m}{\partial x}(x)$ in \eqref{eq4}.  For analysis purpose, assume without loss of generality that $R_Q$ and $R_m$ have the same dimension.

Using the experimental data \eqref{dataset} to offset the lack of information, Theorem 1 has shown that the problem of making the closed-loop system contractive is feasible  if a solution to the data-based SDP \eqref{eq:SDP} exists. 
It can be shown 
that the latter holds if (i) \eqref{eq3} is feasible; (ii) the data are sufficiently rich, namely we can parametrize through data at least one controller that solves the problem with complete information, or, formally, 
\begin{equation} \label{eq6}
\begin{bmatrix} \begin{bmatrix} K_m & 0_{m \times e} \end{bmatrix} \\ I_s \end{bmatrix} \subseteq \text{im} \begin{bmatrix} U_0 \\ Z_0 \end{bmatrix} \text{ for some $K_m \in \mathcal{K}_m$;}
\end{equation}
(iii) $R_Q$ is sufficiently close to $R_m$, that is, there exists a sufficiently small $\varepsilon >0$ such that $\|R_Q-R_m\|\le \varepsilon$. 
{\color{black} To prove this claim,
let $(P_m,K_{m1},K_{m2},\beta)$ be a solution to \eqref{eq3} with $K_m$ satisfying \eqref{eq6}. Thus, there exist matrices $G_1$ and $G_2$ such that
\[
\begin{bmatrix} \begin{bmatrix} K_m & 0_{m \times e} \end{bmatrix} \\ I_n \end{bmatrix} = \begin{bmatrix} U_0 \\ Z_0 \end{bmatrix} \begin{bmatrix} G_1 & G_2 \end{bmatrix}. 
\]
Thus, \eqref{eq:SDP1} and \eqref{eq:SDP4} hold. Moreover,
$\Phi_1 = X_1 G_1$ and $\Phi_2 = X_1 G_2$, hence
\[
\ba{l}
X_1 G_1 P_m + P_m (X_1 G_1)^\top + 
\beta I_n \\ \hspace{2cm} 
+  P_m R_m R_m^\top P_m + X_1 G_2  (X_1 G_2)^\top \preceq 0,
\ea
\]
which implies 
\[
\ba{l}
X_1 G_1 P_m + P_m (X_1 G_1)^\top + \alpha I_n \\ \hspace{2cm} +  P_m R_m R_m^\top P_m + X_1 G_2  (X_1 G_2)^\top \prec 0
\ea
\]
for all $\alpha \in (0,  \beta)$. 
In turn, this implies that there exists a sufficiently small scalar $\varepsilon >0$ such that
\[
\ba{l}
X_1 G_1 P_m + P_m (X_1 G_1)^\top + \alpha I_n  
+  P_m R_Q R_Q^\top P_m \\ + X_1 G_2  (X_1 G_2)^\top \preceq 0  
\text{ for all $R_Q$ such that $\|R_Q-R_m\| \leq \varepsilon$},
\ea
\]
which is a sufficient condition for \eqref{eq:SDP} to hold with $P=P_m$.

}

To summarize, assuming that the model-based problem is feasible also the data-based problem is feasible provided that the data are sufficiently rich in the sense of \eqref{eq6} and that we are able to choose a function $Z$ sufficiently close to $Z_m$, which can be done either by exploiting prior information on the system or through identification techniques. 

We briefly comment on the 
condition \eqref{eq3}. The existence of a tuple $(P_m,K_{m1},K_{m2},\beta)$ such that \eqref{eq3} holds is equivalent to the existence of a triplet $(P_m,K_{m1},K_{m2})$ such that 
\begin{equation} \label{eq3-strict}
\ba{l}
(A_{m1}+B K_{m1}) P_m + P_m (A_{m1}+B K_{m1}) ^\top +\\ 
  P_m R_m R_m^\top P_m + (A_{m2}+B K_{m2})  (A_{m2}+B K_{m2}) ^\top \prec 0.
\ea\end{equation}
The existence of a solution to the Riccati inequality above is equivalent to the problem of making the 
fictitious system 
\[
\dot z = A_{m1} z + Bv + A_{m2} w, \; y= R_m^\top z,
\]
where $v$ is the control input, $w$ the ``disturbance" and $y$ the ``regulated output'', asymptotically stable and its $\mathcal{L}_2$-gain  strictly less than $1$ by the full-information feedback $v=K_{m1} z+ K_{m2} w$. This is a variation of a classical disturbance attenuation problem where both the dynamical matrix $A_{m1}$ and the matrix $A_{m2}$
are modified by feedback. As such, the feasibility of the inequality 
\eqref{eq3} is a well-studied topic within the robust control literature. 

The same arguments can be used to discuss the feasibility of the other SDPs appearing in the paper, including the SDP \eqref{eq:SDP-noise} below, which corresponds to the case of noisy data. In this case, the feasibility of the data-based condition can be still reduced to a model-based Riccati inequality, but the arguments are slightly more involved and are not added here for the sake of conciseness.

%
\medskip

{\color{black} \subsubsection*{ On asymptotic properties} Having enforced the contraction property on the closed-loop system {\color{black}by Theorem \ref{thm:contractivity}}, {\color{black} we discuss} its asymptotic properties, provided an equilibrium exists. {\color{black}The results below are {\color{black}a consequence of} well known {\color{black} properties of contractive systems} and we give them because {\color{black}they are used} later.} 

We recall the definition of region of attraction used below.
\begin{definition}\label{def.ROA} (\cite[Def. 1]{dprt2023cancellation})
Let $\overline{x}$ be an asymptotically stable
equilibrium point for the system $\dot x = f(x)$. A set $\mathcal{R}$ defines a region of attraction (ROA) for the system relative to $\overline{x}$ if for every $x(0) \in \mathcal{R}$ we have $\lim_{t \rightarrow \infty} x(t)=\overline{x}$. 
\end{definition}

%
 
\begin{corollary}\label{cor1}
Let the conditions of Theorem \ref{thm:contractivity} hold and assume additionally that $Z(x_*)=0$ and $\mathcal{X}$ is a convex set containing $x_*$ in its interior. 
Then $x_*$ is the unique\footnote{In fact, it is the unique solution  defined and bounded for all $t\in {\color{black}\R}$ and contained in $\mathcal{X}$.} equilibrium point in $\mathcal{X}$, it is uniformly exponentially stable for $\dot x = (A+BK) Z(x)$ and 
any solution initialized in the set $\mathcal{V}:=\left\{x\in \mathbb{R}^n\colon (x-x_*)^\top {\color{black}P}^{-1} (x-x_*)\le \delta\right\}$, with $\delta>  0$ such that $\mathcal{V}$ is contained in $\mathcal{X}$, converges to $x_*$. Moreover, $\mathcal{V}$ gives an estimate of the ROA relative to $x_*$.
\end{corollary}
 
\emph{Proof.} It is a consequence of Corollary \ref{cor:convergent} in the Appendix. 
\qedp

 \smallskip 
 
The assumption $Z(x_*)=0$ necessarily implies that $x_*=0$, {\color{black} as $Z(x_*)=\begin{bmatrix} x_*^\top & Q(x_*)^\top \end{bmatrix}^\top$.} The assumption can be lifted under a more stringent condition on $\mathcal{X}$. 
 
 \begin{corollary}\label{cor2}
Let the conditions of Theorem \ref{thm:contractivity} hold where $\mathcal{X}$  is either  (i) a convex, closed,  
forward invariant set for 
{\color{black}\eqref{eq:closed}}
 and 
 {\color{black}\eqref{eq:closed}}
is forward complete on $\mathcal{X}$, 
or {\color{black}(ii)} $\mathcal{X}=\mathbb{R}^n$. 
Then there exists a unique equilibrium point $x_*$ in $\mathcal{X}$ for 
{\color{black}\eqref{eq:closed}.}
Moreover, $x_*$  is uniformly exponentially stable for 
{\color{black}\eqref{eq:closed}}
and any solution of 
{\color{black}\eqref{eq:closed}}
initialized in $\mathcal{X}$ uniformly exponentially converges to $x_*$.
\end{corollary}

\smallskip

\emph{Proof.}  
{\color{black}If $\mathcal{X}$ satisfies (i),} then the thesis follows from Theorem \ref{lem:sontag} in the Appendix. If  $\mathcal{X}$ satisfies (ii), then the thesis follows from Theorem \ref{th:demid-pavlov}  in the Appendix. \qedp

%

\subsection{Comparison with other methods and a numerical example}\label{subsec:comparison}

It is useful to compare this approach with other methods  on controller design that are based on convex programming, see \cite{dprt2023cancellation,dpt2023arc,martin2023guarantees}. 
Starting from the data-based representation of the closed-loop dynamics
$\dot x = X_1G_1 x + X_1 G_2 Q(x)$, and by assuming that the term $Q(x)$ acts as a ``remainder" term, i.e. $\lim_{x \rightarrow 0} \frac{|Q(x)|}{|x|}=0$, \footnote{This implies that $\overline x = 0$ is a known equilibrium of the system.} 
a simple approach is to search for matrices 
${\color{black} P \in \mathbb S^{n \times n}, P \succ 0}, Y_1 \in \mathbb{R}^{T \times n}$, and a positive scalar $\alpha$ satisfying the following conditions: 
\begin{subequations}
\label{SDP-taylor}
\begin{alignat}{3}
&\text{\eqref{eq:SDP0}, \eqref{eq:SDP1}, \eqref{eq:SDP4}} \label{SDP-taylor2}, \\
&X_1Y_1 +(X_1Y_1)^\top + \alpha I_n \preceq 0.
\label{SDP-taylor3} 
\end{alignat}
\end{subequations} 
If \eqref{SDP-taylor} is feasible then $u=Kx$ with $K=U_0Y_1{\color{black}P}^{-1}$ locally stabilizes the origin of the closed-loop system. This approach is inspired by Lyapunov's indirect method, which guarantees stability of the nonlinear dynamics by stabilizing the linearized dynamics around the desired equilibrium point. 

Building on \eqref{SDP-taylor}, an alternative approach is to stabilize the linearized dynamics
while simultaneously reducing the impact of the nonlinearities in closed-loop. This amounts to searching for matrices 
${\color{black} P \in \mathbb S^{n \times n}, P \succ 0}, Y_1 \in \mathbb{R}^{T \times n}$, 
$G_2 \in \mathbb{R}^{T \times (s-n)}$ and a positive scalar $\alpha$ satisfying the following program: 
\begin{subequations}
\label{SDP-ncancellation}
\begin{alignat}{3}
\text{minimize}~&  \|X_1G_2\| \label{SDP-ncancellation1}\\
\text{subject to}~&\text{\eqref{eq:SDP0}, \eqref{eq:SDP1}, \eqref{eq:SDP4}}, \eqref{SDP-taylor3}. \label{SDP-ncancellation2}
\end{alignat}
\end{subequations}  
If the minimum is attained at zero, then 
$(A+BK)\frac{\partial Z(x)}{\partial x}= X_1 G_1$  and condition \eqref{contractivity} holds with $\mathcal{X}=\mathbb{R}^n$. In particular, in this case the closed-loop dynamics becomes linear.

Compared with the method proposed in this paper, \eqref{SDP-taylor} and \eqref{SDP-ncancellation} entail milder feasibility conditions since condition \eqref{SDP-taylor3} is less demanding than \eqref{eq:SDP2}. Furthermore, Lyapunov's indirect method allows {\color{black} the designer} to treat the term $X_1 G_2 Q(x)$
that appears in the expression $\dot x = X_1G_1 x + X_1 G_2 Q(x)$ as an \emph{unknown} quantity,
see \cite{dpt2023arc} for details. This is somewhat harder with the method proposed in this paper since the control strategy relies on the full nonlinear behaviour to enforce a stable (contracting) behaviour. On the other hand, the proposed method possesses some nice features, notably:
\begin{enumerate}
\item[(1)] \emph{Stability properties.} The proposed method permits to have \emph{global} stability properties, as long as condition \eqref{contractivity} holds with $\mathcal{X}=\mathbb{R}^n$. This is instead not achievable (at least not by design) with Lyapunov's indirect method, and requires exact nonlineary cancellation, i.e. $X_1G_2=0$, with \eqref{SDP-ncancellation}, which is possible only when the system to control has a specific structure, see \cite{dprt2023cancellation,gdpt2023cdc}.  
\item[(2)]\emph{Knowledge of the equilibrium}. As detailed in Corollary \ref{cor2}, there are case of practical interest in which the proposed method does not require the knowledge of the equilibrium point, which follows because contractive systems  have the incremental stability property that all the trajectories converge to a unique solution, see Theorems \ref{lem:sontag} and \ref{th:demid-pavlov} in the Appendix. In contrast, neither \eqref{SDP-taylor} nor \eqref{SDP-ncancellation} possess this property. For these methods, \emph{exact} knowledge of the equilibrium point is needed to transform the problem into the standard form of a zero-regulation problem. (This is not surprising considering that both \eqref{SDP-taylor} and \eqref{SDP-ncancellation} rest on Lyapunov's indirect method.)
\end{enumerate}
 

We illustrate the foregoing considerations via a numerical example. 

\begin{example}\label{ex1}
Consider the dynamics of a single link manipulator with a flexible joint \cite[Section 4.10]{isidori1995nonlinear}
\begin{subequations}\label{robot.arm}   
\begin{align}
&\dot x_1 = x_2
\\
&\dot x_2 = -\frac{K_c}{J_2}x_1  +\frac{K_c}{J_2 N_c}x_3 - \frac{mgd}{J_2}\cos x_1
\\
&\dot x_3 = x_4
\\
&\dot x_4 = \frac{K_c}{J_1 N_c}x_1 - \frac{K_c}{J_1 N_c^2}x_3  + \frac{1}{J_1}u
\end{align}
\end{subequations}
where $x_1$, $x_3$ denote the angular positions of the link and of the actuator shaft, respectively, and $u$ is the torque produced at the actuator axis. Compared with \cite{isidori1995nonlinear}, to test Theorem \ref{thm:contractivity} the terms due to friction are neglected, as in \cite{marino1986nonlinear}, \cite[Section 5]{spong1987modeling}. We will collect data from the system setting the parameters as $K_c=0.4$, $J_2=0.2$, $N_c=2$, $J_1=0.15$, $m=0.4$, $g=9.8$ and $d=0.1$.

We introduce $Q(x)=  \cos x_1 $. Note that 
\[
\frac{\partial Q}{\partial x}(x)^\top \frac{\partial Q}{\partial x}(x) = {\color{black}\text{diag}(\sin(x_1)^2, 0, 0, 0)}.
\] 
If we set $\mathcal{X} = \mathbb{R}^4$, then \eqref{asspt} is satisfied with 
\be\label{exmpl1:RQ}
R_Q = R_Q ^\top ={\color{black}\text{diag}(1, 0, 0, 0)}.
\ee
We collect $T =10$ samples by running an experiment with input uniformly distributed in  $[-0.1,0.1]$,
and with an initial state within the same interval. The SDP \eqref{eq:SDP} is feasible and returns the controller $K$ {\color{black}and the closed-loop dynamic in \eqref{con.close}}.
\begin{subequations}
\label{con.close}
\begin{align}
 K &=
 \begin{bmatrix}  -3.9338 & -15.3213 &  -7.9442  & -1.8399  &  0.0147 \end{bmatrix}
 \\
 {\color{black}\dot x }& {\color{black}= {\scriptsize \begin{bmatrix}          -0.0000  &  1.0000&    0.0000 &   0.0000 &   0.0000\\
   -2.0000 &  -0.0000 &   1.0000 &   -0.0000 &  -1.9600\\
    0.0000 &   0.0000  &  0.0000  &  1.0000 &   -0.0000\\
  -24.8923 & -102.1423 & -53.6282  & -12.2658   &  0.0982\end{bmatrix} Z(x).}\label{close}}
\end{align}
\end{subequations}
Since $\mathcal{X} = \mathbb{R}^4$, 
by Corollary \ref{cor2}, the equilibrium point $x_* = \begin{bmatrix}  -0.6382 &  0 &  0.2977  & 0 \end{bmatrix}^\top$ obtained from the simulation experiment or analytically from {\color{black} \eqref{close}} is unique  in $\mathbb{R}^4$ for {\color{black}\eqref{close}}. 
Moreover, $x_*$  is globally uniformly exponentially stable for {\color{black} \eqref{close}}. 
\begin{figure*}
\begin{subequations}
\label{con.close.more}
\begin{align}
 &K = \begin{bmatrix}  -160.5762 & -180.5945 & -47.7729 &  -5.2879 &   0.0400 &  -0.0000  &  0.0000 \end{bmatrix} \label{close.more.con}\\ 
& {\color{black}\dot x} = {\color{black}\begin{bmatrix}     -0.0000 &   1.0000 &   -0.0000 &   -0.0000 &   -0.0000 &  0.0000  & 0.0000\\
   -2.0000  & 0.0000 &   1.0000  &  0.0000 &  -1.9600 &   0.0000 &   0.0000\\
    -0.0000 &   -0.0000  &  -0.0000  &  1.0000&   0.0000&    -0.0000  &  -0.0000\\
 -1069.1743 &-1203.9634 &-319.1524  &-35.2523&    0.2666  &  -0.0001  &  0.0000\end{bmatrix} Z(x) }\label{close.more}
\end{align}
\hrulefill
\end{subequations}
\end{figure*}

Next, in order to test the effectiveness of Theorem \ref{thm:contractivity} for other choices of $Q(x)$, we add more nonlinearities in $Q(x)$ and let $Q(x)=  \begin{bmatrix} \cos x_1 & x_1^2 & \sin x_2 \end{bmatrix}^\top$. Note that 
\[
\frac{\partial Q}{\partial x}(x)^\top \frac{\partial Q}{\partial x}(x) = {\color{black}\text{diag}(\sin(x_1)^2 + 4x_1^2, \cos(x_2)^2, 0, 0)}.
\] 
If we set $\mathcal{X} = [-w, w]\times \mathbb{R}^3$, where $w\in \mathbb{R}_{> 0}$, then \eqref{asspt} is satisfied with 
\[
R_Q = R_Q ^\top ={\color{black}\text{diag}(\sqrt{4w^2 +1}, 1, 0, 0)}.
\] 
Here, we set $w = 1$.
We collect $T =10$ samples under the same experiment setup. The SDP \eqref{eq:SDP} is feasible and returns the controller $K$ and {\color{black}the closed-loop dynamics in \eqref{con.close.more}}. 
The unique equilibrium point obtained from the simulation experiment or analytically from {\color{black}\eqref{close.more}} is $x_* = \begin{bmatrix}  -0.3447 &  0 &  1.1554  & 0 \end{bmatrix}^\top$. Note that $x_* \in \text{int}(\mathcal{X})$.
By Corollary \ref{cor:convergent} in the Appendix, any solution of {\color{black}\eqref{close.more}} initialized in any sub-level set of 
$V(x)=(x-x_*)^\top {\color{black}P}^{-1}  (x-x_*)$ contained in  $\mathcal{X}$ uniformly exponentially converges to $x_*$. We observe the following: (i) from numerical results, the solutions of {\color{black}\eqref{close.more}} initialized at the points outside the largest Lyapunov sublevel set contained in  $\mathcal{X}$ still converge to $x_*$, which means the exact ROA is much larger and the obtained controller {\color{black}\eqref{close.more.con}} is possibly a global controller. (ii) \eqref{eq:SDP} remains feasible up to $w = 100$, and as $w$ increases, the magnitude of the coefficients for the linear part of the obtained controller tends to increase, which is consistent with the comments about condition \eqref{eq:SDP}. With $w = 100$, the magnitude is of order $10^5$, which results in a high-gain controller. 

We do not  compare \eqref{eq:SDP} with \eqref{SDP-ncancellation},  since the lack of a specific structure of the systems does not allow \eqref{SDP-ncancellation} to exactly cancel the nonlinearity.  
It is more interesting to compare 
\eqref{eq:SDP}  with the 
{\color{black}data-based implementation of the stabilization by the first-order approximation 
\cite[Theorem 2]{dpt2023arc}.
}  
{\color{black}
Denote system \eqref{robot.arm} as $\dot {x} = f(x,u)$
and assume that the desired equilibrium point $(x_*, u_*)$ is known, i.e. $x_* = \begin{bmatrix}  -0.3447 &  0 &  1.1554  & 0 \end{bmatrix}^\top$ and $u_* = 0.1845$.   
Linearize system \eqref{robot.arm} at  $(x_*, u_*)$ 
which yields
\[
\dot {\tilde x} = A {\tilde x} + B {\tilde u} + r(\tilde x,\tilde u)
\]
where $\tilde x = x - x_*$,  $\tilde u = u - u_*$ are the shifted state and input variables, 
\[
A = \left.\displaystyle \frac{\partial f}{\partial {x}}\right |_{(x, u)= (x_*, u_*)}, \quad
B = \left.\displaystyle \frac{\partial f}{\partial {u}}\right |_{(x, u)= (x_*, u_*)}
\]
are unknown constant matrices
and $r\colon \mathbb R^4\times \mathbb{R}\to \mathbb R^4$
denotes the unknown remainder. Collect 
$\{( x_i, u_i, \dot{x}_i)\}_{i=0}^{T-1}$ from system \eqref{robot.arm} and compute the samples
$\{(\tilde x_i, \tilde u_i, \dot{\tilde x}_i)\}_{i=0}^{T-1}$, which are used to define the data matrices 
\begin{subequations}
\label{eq:data.shift}
\begin{align}
&\tilde U_0 :=  \begin{bmatrix}
\tilde u_0& \tilde u_1 & \ldots & \tilde u_{T-1}
\end{bmatrix}\in \R^{m\times T}\,, \\[0.2cm]
&\tilde X_0 :=   \begin{bmatrix} \tilde x_0 & \tilde x_1 & \ldots & \tilde x_{T-1} 
\end{bmatrix}\in \R^{n\times T}\,,  \\[0.2cm]
&\tilde X_1 :=    \begin{bmatrix} \dot{\tilde x}_0 & \dot{\tilde x}_1 & \ldots & \dot{\tilde x}_{T-1} 
\end{bmatrix}\in \R^{n\times T}\,.
\end{align}
\end{subequations}
These matrices 
satisfy the identity
$\tilde X_1 = A\tilde  X_0 + B\tilde U_0 + R_0$,
where 
$R_0 := \begin{bmatrix} r_0 & r_1 & \cdots & r_{T-1}  \end{bmatrix} \in \mathbb R^{n \times T}$
is the (unknown) data matrix of samples of the reminder $r$.
Under the control law $\tilde u = K \tilde x$, the data-based representation of the closed-loop dynamics is 
\begin{equation} \label{eq:reminder}
\dot {\tilde x} = (\tilde X_1 - R_0)G {\tilde x} + r(\tilde x,\tilde u).
\end{equation}
Let $\mathcal{R}:=\{
R\in \mathbb{R}^{n\times T}\colon RR^\top \preceq \Delta \Delta^\top\}$ where $\Delta \in \mathbb{R}^{n\times p}$ is chosen by the designer, and assume $R_0\in \mathcal{R}$. Similar to \eqref{SDP-taylor}, a simple approach to obtain a stabilizing controller $K$ is to search for matrices 
${\color{black} P \in \mathbb S^{n \times n}, P \succ 0}, Y_1 \in \mathbb{R}^{T \times n}$, and positive scalars $\alpha$ and $\mu$ satisfying the following program \cite{dpt2023arc}:  
\begin{subequations}
\label{SDP-taylor.reminder}
\begin{alignat}{3}
&\text{\eqref{eq:SDP0}, \eqref{eq:SDP1}, \eqref{eq:SDP4}}, \\
&\begin{bmatrix} X_1 Y_1 +(X_1 Y_1)^\top +  \alpha I_n + \mu \Delta \Delta^\top &{\color{black}\star}  \\ 
Y_1  & -\mu I_T
\end{bmatrix} \preceq 0. 
\end{alignat}
\end{subequations} 
If SDP \eqref{SDP-taylor.reminder} with $X_1$ replaced by $\tilde X_1$  is feasible then $\tilde u = K \tilde x$ with $K=\tilde U_0Y_1 {\color{black}P} ^{-1}$ locally stabilizes the origin of the closed-loop system \eqref{eq:reminder}. We collect $T =10$ samples from system \eqref{robot.arm}. In this example, the reminder term is $r(\tilde x,\tilde u) = \begin{bmatrix}  0 &  1 &  0  & 0 \end{bmatrix}^\top \overline{r}(\tilde x)$ with $\overline{r}(\tilde x) :=  - \frac{mgd \cos x_{*1}}{J_2} (\cos {\tilde x}_1 - 1) + \frac{mgd \sin x_{*1}}{J_2} (\sin {\tilde x}_1 - {\tilde x}_1)$. Let $\delta(\tilde x) := 4|\cos {\tilde x}_1 - 1| + 2|\sin {\tilde x}_1 - {\tilde x}_1|$, which over-approximates $\overline{r}(\tilde x)$ by more than 100\%. Then, we set $\Delta = \sqrt{T} \text{diag}(0,c,0,0)$, with $c$  equal to the maximum of $\delta(\tilde x)$ over the experimental data, such that $R_0\in \mathcal{R}$ is satisfied. 
{\color{black} Once we replace  $\sqrt{T} c$  with its numerical value, then we obtain $\Delta = \text{diag}(0,0.0197,0,0)$.}
Then the SDP \eqref{SDP-taylor.reminder} is feasible and returns the controller $K = {\color{black}\begin{bmatrix}  11.7829 &  -5.9117 & -18.9576  & -1.8546 \end{bmatrix}}$. 
For this controller, we numerically determine the set $\mathcal{H} = \{ \tilde x: \dot V(\tilde x) <0\}$, with $\dot V(\tilde x):=2 \tilde x^\top {\color{black}P}^{-1} \dot{\tilde x}$, over which the Lyapunov function $V(\tilde  x) = \tilde x^\top {\color{black}P}^{-1} \tilde x$ decreases and any sub-level set $\mathcal{R}_\gamma := \{ \tilde x: V(\tilde x) \leq \gamma\}$ of $V$ with $\gamma >0$ contained in $\mathcal{H} \cup \{0\}$ gives an estimate of the ROA for the closed-loop system. Planar projections of the set $\mathcal{H}$ and of a sub-level set of $V$ are displayed in Figure \ref{ROA.taylor}.  Hence, the controller $u = K(x- x_*)+u_*$ renders $(x_*, u_*)$ a locally asymptotically stable equilibrium point for system \eqref{robot.arm}.
\begin{figure}[ht!]
\includegraphics[scale=0.45]{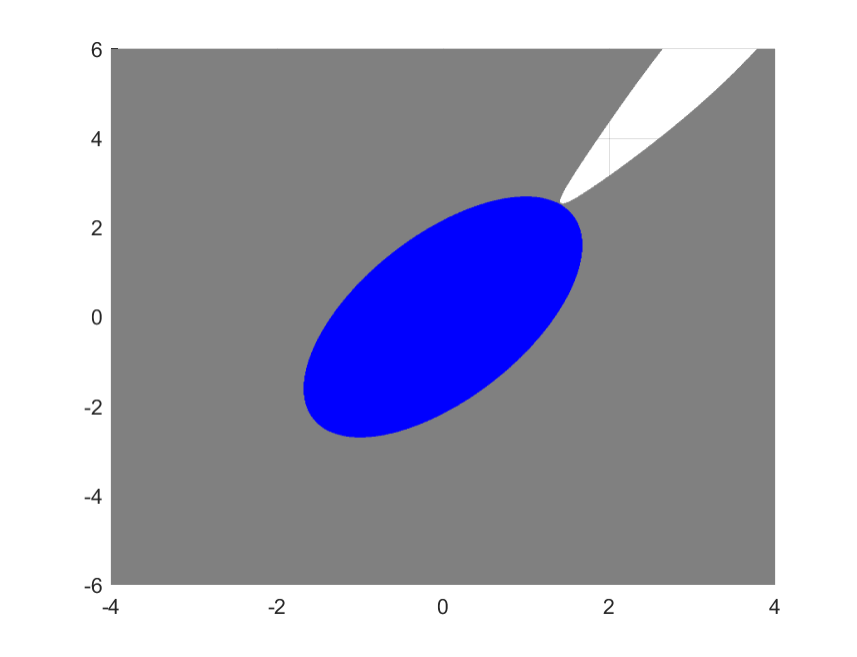}
   \caption{The grey
set represents the projection of the set $\mathcal{H}$
onto the plane $\{ \tilde x: \tilde x_3=\tilde x_4=0\}$. The blue set is the projection onto the same plane of a Lyapunov sub-level set $\mathcal{R}_\gamma$ contained in $\mathcal{H}$; here, {${\color{black}P}^{-1} = \left[{\color{black}\begin{smallmatrix}
    8.4854 &  -3.1359 &  -9.8650 &  -0.5898\\
   -3.1359 &   3.2924  &  5.7813  &  0.2450\\
   -9.8650  &  5.7813 &  16.4525  &  0.8604\\
   -0.5898 &   0.2450 &   0.8604 &   0.1376\\
 \end{smallmatrix}}\right]$} and $\gamma = 15.6$. }\label{ROA.taylor}
\end{figure}
}

{\color{black}To further show the effectiveness of the proposed control design, we test it on another well-known  nonlinear control benchmark, i.e. the surge system \cite{krstic1995nonlinear}. 
}
\end{example}

\begin{example}
{\color{black}Consider the surge subsystem of an axial compressor model 
\begin{subequations}\label{surge}
\begin{align}
&\dot x_1 = -x_2 - \frac{3}{2} x_1^2 - \frac{1}{2} x_1^3,
\nonumber
\\
&\dot x_2 =  {\color{black}u}, \nonumber
\end{align}
\end{subequations}
where $x_1$ and $x_2$ denote the  deviations of the flow and 
pressure from their respective set points, and $u$ is the control input.  Note that the origin $x_*=0$ is an unstable equilibrium of the uncontrolled system. The control objective is to make $x_*$ an asymptotically stable equilibrium for the closed-loop system and we use contractivity to achieve the goal. 
We introduce $Q(x)=  \begin{bmatrix} x_1^2 & x_1^3 \end{bmatrix}^\top $. Note that 
\[
\frac{\partial Q}{\partial x}(x)^\top \frac{\partial Q}{\partial x}(x) = \text{diag}(4x_1^2 + 9x_1^4, 0).
\] 
If we set $\mathcal{X} = [-w, w]\times \mathbb{R}$, where $w\in \mathbb{R}_{> 0}$, then \eqref{asspt} is satisfied with 
\[
R_Q = R_Q ^\top =\text{diag}(\sqrt{4w^2 + 9w^4}, 0).
\]
Here, we set $w = 1$. We collect $T = 10$ samples running an experiment with input uniformly distributed in  $[-1,1]$,
and with an initial state within the same interval. The SDP is feasible and returns the controller {\color{black}$K= \begin{bmatrix} 472.8008 & -26.7351 &   0.9875  &  0.1960 \end{bmatrix}$}.
Note that $Z(x_*) = 0$ and $x_* \in \text{int}(\mathcal{X})$, and by Corollary 1, any solution of the closed-loop system initialized in any sub-level set of 
$V(x)=x^\top P^{-1}  x$ contained in  $\mathcal{X}$ uniformly exponentially converges to the origin. {\color{black}For  controller $K$, we numerically determine the set $\mathcal{H} = \{ x: \dot V(x) <0\}$, with $\dot V(x):=2 x^\top P^{-1} \dot{x}$, over which the Lyapunov function $V(x) =  x^\top P^{-1} x$ decreases and any sub-level set $\mathcal{R}_\gamma := \{ x: V(x) \leq \gamma\}$ of $V$ with $\gamma >0$ contained in $\mathcal{H} \cup \{0\}$ gives an estimate of the ROA for the closed-loop system. The set $\mathcal{H}$ and a sublevel set of $V$ are shown in Figure \ref{surge-ROA1}. Note that the blue set is larger than any sub-level set of 
$V$ contained in  $\mathcal{X}$. In fact, providing stability certificate imposes some conservativeness on the controller design procedure, and the resulting controller typically performs better than theoretically predicted.} 
\begin{figure}[ht!] \centering
\includegraphics[scale=0.45]{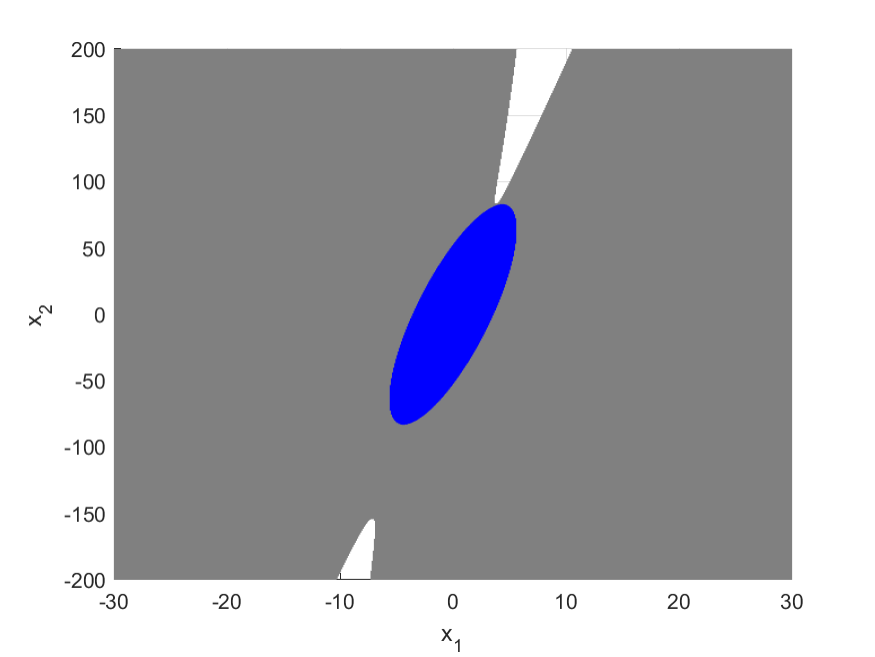}
   \caption{The grey
set represents the set $\mathcal{H}$, while the blue set is a Lyapunov sub-level set $\mathcal{R}_\gamma$ contained in $\mathcal{H}$; here, $P^{-1} = \begin{bmatrix}
    7.1505  & -0.3761 \\
   -0.3761  &  0.0335
\end{bmatrix}$ and $\gamma = 95$.}\label{surge-ROA1}
\end{figure}

Next, we add more nonlinearities in $Q(x)$ and let $Q(x)=  \begin{bmatrix} x_1^2 & x_1^3 & x_2^2 & x_2^3  \end{bmatrix}^\top$. Note that 
\[
\frac{\partial Q}{\partial x}(x)^\top \frac{\partial Q}{\partial x}(x) = \text{diag}(4x_1^2 + 9x_1^4, 4x_2^2 + 9x_2^4).
\] 
If we set $\mathcal{X} = [-w_1, w_1]\times [-w_2, w_2]$, where $w_1, w_2\in \mathbb{R}_{> 0}$, then $\frac{\partial Q}{\partial x}(x)^\top \frac{\partial Q}{\partial x}(x)\preceq R_Q R_Q^\top$ is satisfied with 
\[
R_Q = R_Q ^\top =\text{diag}(\sqrt{4w_1^2 + 9w_1^4}, \sqrt{4w_2^2 + 9w_2^4}).
\]
Here, we set $w_1 = 1$ and $w_2 = 0.1$. We collect $T =10$ samples under the same experiment setup. The SDP is feasible and returns the controller $K= \begin{bmatrix} 418.8709 & -37.8660  &  0.5014  &  0.3406  & -0.0002   & 0.0001
 \end{bmatrix}$. 
 Analogously, for this controller, we numerically determine the set $\mathcal{H}$ and a sublevel set of $V$ contained in $\mathcal{H}$ as an estimate of the ROA, which are shown in Figure \ref{surge-ROA2}.

 \begin{figure}[ht!] \centering
\includegraphics[scale=0.45]{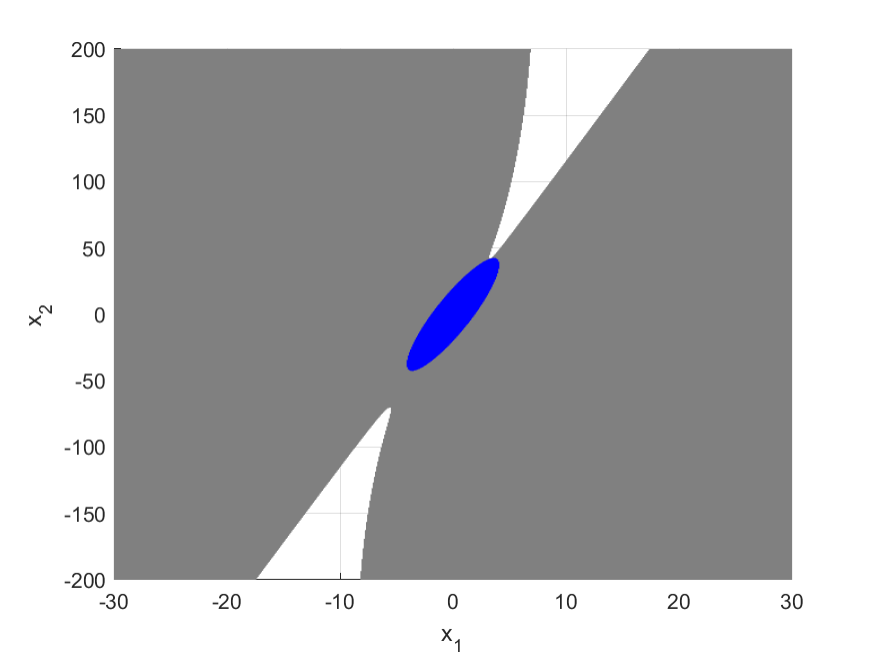}
   \caption{The grey
set represents the set $\mathcal{H}$, while the blue set is a Lyapunov sub-level set $\mathcal{R}_\gamma$ contained in $\mathcal{H}$; here, $P^{-1} = \begin{bmatrix}
   18.8524  & -1.6006 \\
   -1.6006 &   0.1770
\end{bmatrix}$ and $\gamma = 75$.}\label{surge-ROA2}
\end{figure}

}

\end{example}

\smallskip


%
%
%
%
%


{\color{black}

\section{ An extension to general nonlinear systems}
\label{subsec:State-dependent input}

The results of the paper can be extended to a more general class of nonlinear systems, which contains as a special case not only the system \eqref{system2} but also the important class of input-affine systems. In fact, here we consider systems of the form 
\begin{equation}
\label{system.g(x).more.gen}
\dot x = f(x,u).
\end{equation}
We introduce the stacked vector of states and inputs   
$\xi :=[\begin{smallmatrix} x \\ u \end{smallmatrix}]$, 
and consistently with the previous section we assume the following:
\begin{assumption} \label{ass:Z-W.more.gen}
A continuously differentiable function $\mathcal{Z}: \mathbb R^{n+m} \rightarrow \mathbb R^R$ is known such that
$f(x,u)=A\mathcal{Z}(\xi)$ for some matrix $A \in \mathbb R^{n \times R}$. \qedd
\end{assumption} 

Under this assumption the system dynamics \eqref{system.g(x).more.gen} is written as 
\begin{equation}
\label{system.g(x).more.gen-with-A}
\dot x = 
A \mathcal{Z}(\xi).
\end{equation}
We partition 
$\mathcal{Z}(\xi)$ into subvectors $\xi$ and 
$\mathcal{Q}:\R^{n+m} \to \R^{R-n-m}$, where $\mathcal{Q}(\xi)$ collects all the entries of $\mathcal{Z}(\xi)$  that are nonlinear functions of the extended state $\xi$, namely we consider
\begin{equation}\label{Z(x,u)}
\mathcal{Z}(\xi)= \begin{bmatrix} \xi \\ \mathcal{Q}(\xi) \end{bmatrix}.
\end{equation}
Finally, we regard $\xi$ as the new state variable and extend the dynamics \eqref{system.g(x).more.gen-with-A} with  the integral control $\dot u = v$,
where $v\in \mathbb{R}^m$ is a new control input. Other uses of the integral control will be considered in Section \ref{sec:integral-control}.
We obtain the overall system 
\begin{equation}\label{nonl.new.form.more.gen}
\dot \xi = \mathcal{A} \mathcal{Z}(\xi)+\mathcal{B} v,
\end{equation}
where 
\begin{equation}\label{calA-calB}
\mathcal{A} := \begin{bmatrix} \, \overline A & \hat A \\
{0}_{m\times (n+m)} & {0}_{m\times (R-n-m)} \\ \end{bmatrix}, \quad
\mathcal{B} := \begin{bmatrix} 0_{n\times m} \\ I_m \end{bmatrix}{\color{black},}
\end{equation}
having partitioned $A$ as $A=\begin{bmatrix}\, \overline A & \hat A  \end{bmatrix}$ with 
$\overline A\in \R^{n\times (n+m)}$ {\color{black}and 
$\hat A\in \R^{n\times (R-n-m)}$}. 
We  then aim at the design of the feedback control 
\be\label{extended-controller}
v = \mathcal{K} \mathcal{Z}(\xi){\color{black},} 
\ee
where $\mathcal{K}=\begin{bmatrix}\, \overline{\mathcal{K}}& \hat{\mathcal{K}} \end{bmatrix}$, {\color{black}
$\overline{\mathcal{K}}\in \R^{m\times (n+m)}$ and 
$\hat{\mathcal{K}}\in \R^{m\times (R-n-m)}$,
}
to enforce desired properties on  $\dot \xi = (\mathcal{A}+\mathcal{B}\mathcal{K}) \mathcal{Z}(\xi) $. 
{\color{black} Note that with this partition, we follow the control design strategy in Theorem \ref{thm:contractivity} of  differentiating between the linear and the nonlinear part of the extended system \eqref{nonl.new.form.more.gen} and the controller \eqref{extended-controller} (see the remark {\color{black}``On condition (12)" in Section \ref{subsec:discussion}} 
for an explanation of the design strategy). In this way,}
the design can be pursued as in the previous section {\em mutatis mutandis}. Below we concisely present the main  result. 

Collect the dataset {\color{black}$\{x_i, u_i, v_i, \dot x_i\}_{i=0}^{T-1}$} from the system \eqref{nonl.new.form.more.gen} 
and define the data matrices 
\begin{equation*}
\begin{array}{rl}
V_0 :=  & \hspace{-3mm}   
{\color{black}
\begin{bmatrix}
v_0& v_1 & \ldots & v_{T-1}
\end{bmatrix}
}
\in \R^{m\times T}, \\[0.2cm]
\Xi_0 := & \hspace{-3mm}   
{\color{black}
\begin{bmatrix} \xi_0 & \xi_1 & \ldots & \xi_{T-1}
\end{bmatrix}
}
\in \R^{(n+m)\times T},  \\[0.2cm]
\Xi_1 :=  & \hspace{-3mm}   
{\color{black}
\begin{bmatrix} \dot \xi_1 & \dot{\xi}_2 & \ldots & \dot\xi_{T-1} 
\end{bmatrix}
}
\in \R^{(n+m)\times T}, \\[0.2cm]
 \mathcal{Z}_0 := & 
\hspace{-3mm} 
{\color{black} 
\begin{bmatrix} \xi_0 & \xi_1 & \ldots & \xi_{T-1} \\[0.5mm]
\mathcal{Q}(\xi_0) & \mathcal{Q}(\xi_1) & \ldots & \mathcal{Q}(\xi_{T-1})
\end{bmatrix}
} 
\in 
\R^{R\times T},
\end{array}
\end{equation*}
which satisfy the identity $\Xi_1 = \mathcal{A} \mathcal Z_0 + \mathcal{B} V_0$. {\color{black} Bearing in mind that $\xi :=[\begin{smallmatrix} x \\ u \end{smallmatrix}]$ and the functions in $\mathcal{Q}$ are known, the computation of $\Xi_0$ and  $\mathcal Z_0$ requires the samples 
$\{x_i, u_i\}_{i=0}^{T-1}$. On the other hand, since by design $\dot u=v$, the computation of $\Xi_1$ requires the samples 
$\{v_i, \dot x_i\}_{i=0}^{T-1}$.}

{\color{black}Under Assumption \ref{ass:Z-W.more.gen}, the following result provides a dynamic feedback control that imposes contractivity on the closed-loop system made of the controller and the nonlinear system \eqref{system.g(x).more.gen}, thus generalizing Theorem \ref{thm:contractivity}.}
\begin{theorem}\label{cor:approx.extended.more.gen}
Consider the nonlinear system \eqref{system.g(x).more.gen}. Let Assumption \ref{ass:Z-W.more.gen} hold and $\mathcal{Z}(\xi)$ be of the form \eqref{Z(x,u)}, with $\xi =[\begin{smallmatrix} x \\ u \end{smallmatrix}]$. 
Let 
$\mathcal{R}_Q\in \mathbb R^{(n+m) \times {\color{black}\tau}}$ be a known matrix and $\mathcal{W}\subseteq \mathbb{R}^n\times \mathbb{R}^m$ a set such that
\be\label{asspt-extended}
\frac{\partial \mathcal{Q}}{\partial \xi}(\xi)^\top \frac{\partial \mathcal{Q}}{\partial \xi}(\xi)\preceq \mathcal{R}_{\mathcal{Q}} \mathcal{R}_{\mathcal{Q}}^\top \text{ for any $\xi\in \mathcal{W}$}.
\ee
Consider 
the following SDP in the decision variables ${\color{black}\mathcal{P}} \in \mathbb S^{(n+m) \times (n+m)}$,
$\mathcal{Y}_1 \in \mathbb R^{T \times (n+m)}$, 
$\mathcal{G}_2 \in \mathbb R^{T \times (R-n-m)}$ and $\alpha \in \mathbb R_{> 0}$:
\begin{subequations}
\label{eq:SDP-general}
\begin{alignat}{6}
& {\color{black}\mathcal{P}} \succ 0, \\
& \mathcal{Z}_0 \mathcal{Y}_1 = \begin{bmatrix} {\color{black}\mathcal{P}} \\ 0_{(R-n-m) \times (n+m)} \end{bmatrix} \,, \\
& \begin{bmatrix} \Xi_1 \mathcal{Y}_1 +(\Xi_1 \mathcal{Y}_1)^\top + \alpha I_{n+m}  & {\color{black}\star} & {\color{black}\star}  \\ 
{\color{black}\mathcal{G}_2^\top \Xi_1^\top}  & - I_{R-n-m} & {\color{black}\star} \\
\mathcal{R}_{\mathcal{Q}}^\top {\color{black}\mathcal{P}} & 0 & - I_{{\color{black}\tau}}
\end{bmatrix}  \preceq 0 \,,  \\
& \mathcal{Z}_0 \mathcal{G}_2 = \begin{bmatrix} 0_{(n+m) \times (R-n-m)} \\ I_{R-n-m} \end{bmatrix} \,.
\end{alignat}
\end{subequations} 
If the program is feasible then the dynamic feedback control
\be\label{state-feedback-general}
\dot u=\mathcal{K} \mathcal{Z}(\xi){\color{black}} 
\ee 
with 
\begin{eqnarray} \label{eq:K_SDP-general}
\mathcal{K}= V_0 \begin{bmatrix} \mathcal{Y}_1 {\color{black}\mathcal{P}}^{-1}
& \mathcal{G}_2 \end{bmatrix}{\color{black}}
\end{eqnarray}
are such that  the closed-loop dynamics \eqref{system}, \eqref{state-feedback} are exponentially contractive on $\mathcal{W}$. 
\end{theorem}

{\em Proof.} It is similar to the proof of Theorem \ref{thm:contractivity} and is omitted. \qedp

The asymptotic properties of the closed-loop system can be established similarly to Corollaries \ref{cor1}  and \ref{cor2}. For the sake of brevity, we do not formally state them. 

{\color{black}
Theorem \ref{cor:approx.extended.more.gen} shows how to design controllers that enforce contractivity on the class of nonlinear systems \eqref{system.g(x).more.gen}, which is much more general than \eqref{system}. This can be achieved under Assumption \ref{ass:Z-W.more.gen} (instead of Assumption \ref{ass:Z}), which requires $f(x,u)$ to belong to a finite dimensional space of functions spanned by known basis functions, and the condition \eqref{asspt-extended}, which bounds the growth rate of the functions in the library. The result is achieved by extending the dynamics \eqref{system.g(x).more.gen} via an $m$-dimensional integrator, which reduces the overall system to one with a state independent input matrix, thus allowing us to repeat the analysis in the previous section. The price to pay for this extension is a dynamical controller, instead of a static one, and an SDP of larger size, hence more computationally expensive. Other computationally tractable data-based nonlinear control design techniques that do not go through such an extension apply to affine systems $\dot x = f(x)+g(x)u$, with $f,g$ polynomial vector fields, in which case the design can be reduced to a (more computationally expensive) SOS program without the need of a dynamics extension (cf.~\cite{dai2020semi,guo2021data}). }

\begin{example}
Consider the dynamics of a continuous stirred tank reactor \cite{longchamp1980stable}
\begin{subequations}\label{tank.reactor}   
\begin{align}
&\dot x_1 = 4.25 x_1 + x_2 - 0.25 u - x_1 u,
\\
&\dot x_2 = -6.25 x_1 - 2 x_2
\end{align}
\end{subequations}
where $x_1$ and $x_2$ are the deviations from the steady-state output temperature and concentration, respectively, and the control input $u$ is the effect of coolant flow on the chemical reaction. Note that the open-loop system is unstable.
The control objective is to maintain the output temperature and concentration close to their steady-state values by regulating the flow of coolant fluid.

We introduce $\xi :=[\begin{smallmatrix} x \\  u \end{smallmatrix}]$ and $Q(\xi)=  \xi_1 \xi_3$, and then the overall system is given as 
\begin{equation}\label{overall}
\dot \xi = \mathcal{A} \mathcal{Z}(\xi)+\mathcal{B} v
\end{equation}
where 
\begin{equation}\label{calA-calB.ex}
\mathcal{A} := \begin{bmatrix} 4.25 & 1 & -0.25 & -1  \\
-6.25 & -2 &0 &0\\  0 &0& 0& 0\end{bmatrix}, \quad
\mathcal{B} := \begin{bmatrix} 0\\ 0 \\ 1 \end{bmatrix}.
\end{equation} Note that $\frac{\partial {\color{black}\mathcal{Q}}}{\partial \xi}(\xi)^\top \frac{\partial {\color{black}\mathcal{Q}}}{\partial \xi}(\xi) = \begin{bmatrix} \xi_3^2 & 0& \xi_1\xi_3\\  0 & 0 & 0  
\\\xi_1\xi_3 & 0 & \xi_1^2  \end{bmatrix}$. Let $\mathcal{W}=[-w, w] \times \mathbb R \times [-w, w]$, where $w=0.05$ and let $\mathcal{R}_{\mathcal{Q}}=  \mathcal{R}_{\mathcal{Q}}^\top = {\color{black}\text{diag}(0.1, 0, 0.1)}$ such that \eqref{asspt-extended} is satisfied. 

We collect $T =10$ samples by running an experiment on system \eqref{overall} with input uniformly distributed in  $[-0.1,0.1]$,
and with an initial state within the same interval.
The SDP \eqref{eq:SDP-general} is feasible and returns the following controller $\mathcal{K}$ and closed-loop dynamics \begin{subequations}
\label{con.close-extended}
\begin{align}
 &\mathcal{K} =\begin{bmatrix}  851.2634 & 169.3755 & -27.9383 &  -0.9400\end{bmatrix}\\
 &\dot \xi = {\scriptsize \begin{bmatrix}    4.2500 &   1.0000 &  -0.2500  & -1.0000 \\
   -6.2500  & -2.0000  &  0.0000 &   0.0000 \\
   851.2634 & 169.3755 & -27.9383 &  -0.9400\end{bmatrix}} \mathcal{Z}(\xi) \label{close.general}
\end{align}
\end{subequations}
and by Theorem \ref{cor:approx.extended.more.gen}, \eqref{close.general} is exponentially contractive on $\mathcal{W}$.

\begin{figure}[ht!]
\includegraphics[scale=0.45]{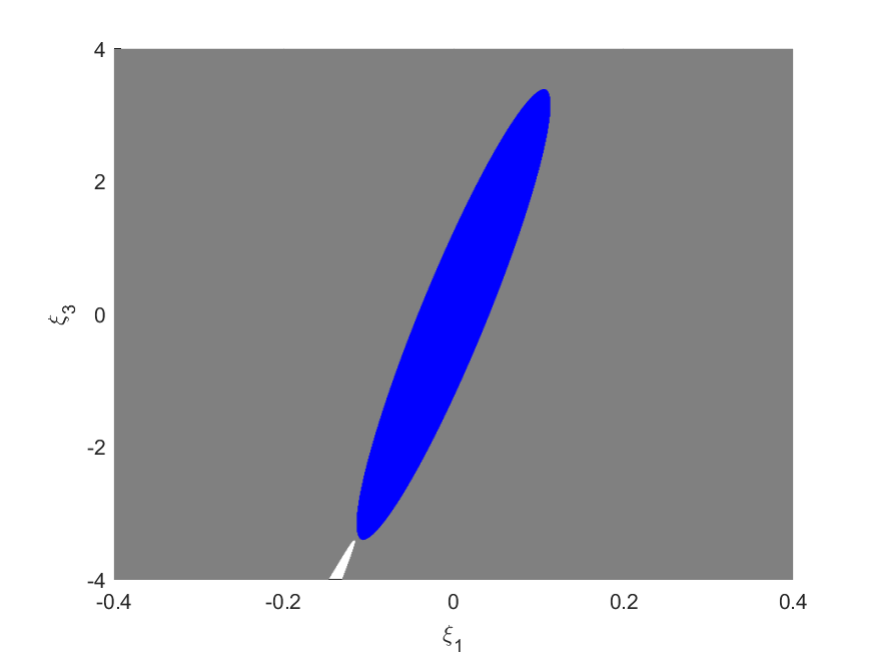}
   \caption{The grey
set represents the projection of the set $\mathcal{H}$
onto the plane $\{\xi: \xi_2=0\}$. The blue set is the projection onto the same plane of a Lyapunov sub-level set $\mathcal{R}_\gamma$ contained in $\mathcal{H}$; here, ${\color{black}\mathcal{P}}^{-1} = \left[\begin{smallmatrix}
   32.9160  &  6.5218 &  -1.0322\\
    6.5218  &  1.3114  & -0.2042\\
   -1.0322  & -0.2042  &  0.0376 \end{smallmatrix}\right]$ and $\gamma = 0.06$.
   }\label{ROA.general}
\end{figure}

Note that $\mathcal{Z}(0) = 0$ and $ \mathcal{W}$ is a convex set containing the origin. By Corollary \ref{cor1}, any solution of \eqref{close.general} initialized in any sub-level set of\footnote{{\color{black}In this case, the Lyapunov function $V$ 
is a function of $\xi :=[\begin{smallmatrix} x \\ u \end{smallmatrix}]$, hence it depends on both the state $x$ and the input $u$. This should not come as a surprise, as the extension of the control system via the $m$-dimensional integrator $\dot u =v$ makes $u$ effectively an additional state variable.}} 
$V(\xi)=\xi^\top {\color{black}\mathcal{P}}^{-1} \xi$ contained in  $\mathcal{W}$ uniformly exponentially converges to the origin. For this controller, we numerically determine the set $\mathcal{H} = \{ \xi: \dot V(\xi) < 0\}$, with $\dot V(\xi):=2 \xi^\top {\color{black}\mathcal{P}}^{-1} \dot \xi$, over which the Lyapunov function $V(\xi) =  \xi^\top {\color{black}\mathcal{P}}^{-1}\xi$ computed along the solutions of \eqref{close.general} decreases and any sub-level set $\mathcal{R}_\gamma := \{ \xi: V(\xi) \leq \gamma\}$ of $V$ with $\gamma >0$ contained in $\mathcal{H} \cup \{0\}$ gives an estimate of the ROA for the closed-loop system \eqref{close.general}. Planar projections of the set $\mathcal{H}$ and of a sub-level set of $V$ are displayed in Figure \ref{ROA.general}.  Note that $\mathcal{W}$ is determined by $w$, i.e. the range of $\xi_1$ and $\xi_3$, and thus we project the set $\mathcal{H}$
onto the plane $\{\xi: \xi_2=0\}$ to observe the Lyapunov sub-level set contained in $\mathcal{W}$. Moreover, we observe that \eqref{eq:SDP-general} remains feasible up to $w=0.1$, and as $w$ increases, the magnitude of the coefficients for the linear part of the obtained controller tends to increase. With $w = 0.1$ and $\mathcal{R}_{\mathcal{Q}}= \mathcal{R}_{\mathcal{Q}}^\top = {\color{black}\text{diag}(0.2, 0, 0.2)}$ such that \eqref{asspt-extended} is satisfied, the magnitude is of order $10^3$ in this case.
\end{example}

{\color{black}In the sections below, we return to consider nonlinear systems with constant input vector fields, which allow us to present the results in a more compact way than what we could do for the general system $\dot x = f(x,u)$ considered here. However, all the results presented next should be applicable to the general case as well.} 

}

\section{Establishing contractivity with noisy data {\color{black}and remainders}}\label{sec:noisy-data-and-remainders}

{\color{black}\subsection{Noisy data}}\label{sec:noisy-data} In this section we extend the previous results to the setup where an additive process disturbance affects the dynamics both during the data acquisition phase and the execution of the control task. Beside the interest per se, establishing contractivity in the presence of disturbance and from noisy data {\color{black}has} interesting implications in the solution of other control tasks than stabilization, as we will investigate later, {\color{black} when we deal with the design of integral controllers for nonlinear systems}. 

In the presence of process disturbances, system \eqref{system} becomes
\begin{equation}
\label{system-with-dist}
\dot x = f(x)+ Bu+Ed 
\end{equation}  
which, under Assumption \ref{ass:Z}, can be written as
\begin{equation}
\label{system2-with-dist}
\dot x = A Z(x) + B u+Ed
\end{equation}  
where $d\in \mathbb{R}^q$ is an unknown time-varying signal representing a process disturbance and $E \in \mathbb{R}^{n\times q}$ is a known matrix that indicates which parts of the dynamics are affected by $d$. {\color{black} If these are not known, then one sets $E=I_n$.  }

Similar to the noiseless case, we perform an experiment to collect the dataset $\mathbb D$ in \eqref{dataset}  and arrange the data into the matrices $U_0, X_0, X_1, Z_0$ defined in \eqref{eq:data}. Note that now  these matrices are related by the equation $X_1=A Z_0+B U_0 +ED_0$, where
\be\label{eq:data5}
D_0 := \begin{bmatrix} d_0 & d_1 & \cdots & d_{T-1}  \end{bmatrix} 
\in \mathbb R^{q \times T} 
\ee
is an unknown matrix. 

A data-dependent representation of system \eqref{system-with-dist} in closed-loop with the controller $u=KZ(x)$ can also be given in the presence of disturbances. In fact, the following holds:
\begin{lem} \label{lem:main_noise}
Consider any matrices $K \in \mathbb R^{m \times s}$,
$G \in \mathbb R^{T \times s}$ such that \eqref{eq:GK} holds. 
Let $G$ be partitioned as $G = \begin{bmatrix} G_1 & G_2 \end{bmatrix}$, where
$G_1 \in \mathbb R^{T \times n}$ and $G_2 \in \mathbb R^{T \times (s-n)}$.
Let Assumption \ref{ass:Z} hold.
Then system \eqref{system-with-dist} under the control law 
\eqref{state-feedback}
results in
the closed-loop dynamics 
\begin{equation} \label{eq:GK_closed_noise}
\!\!\!
\ba{l}
\dot x = (A+BK) Z(x) +Ed \\
= (X_1-E D_0)G_1 x + (X_1-E D_0)G_2  Q(x) +Ed. 
\ea\end{equation}
\end{lem} 

\emph{Proof.} See \cite[Lemma 2]{dprt2023cancellation}. 
\qedp

\smallskip

To further the design and the analysis, we impose some requirements on $D_0$, namely we introduce the following assumption:
\begin{assumption}\label{asspt:noise}
For a given $\Delta$, 
$D_0\in \mathcal{D}=\{
D\in \mathbb{R}^{q\times T}\colon DD^\top \preceq \Delta \Delta^\top
\}$. \qedd
\end{assumption}
This assumption on the disturbance has been shown to be flexible enough to capture various classes of disturbances, such as disturbances satisfying energy and pointwise bounds as well as stochastic disturbances. We refer the  reader to \cite{de2019formulas,bisoffi2022data, BisoffiSCL2021tradeoffs,dprt2023cancellation} for details.

Arguments similar to those used to prove Theorem \ref{thm:contractivity} lead to the following:
\begin{theorem} \label{thm:contractivity-noise}
Consider the nonlinear system  \eqref{system-with-dist}, let Assumption \ref{ass:Z} hold 
and $Z(x)$ be of the form \eqref{eq:Z}. 
Let 
$ R_Q\in \mathbb R^{n \times {\color{black}\tau} }$ be a known matrix and $\mathcal{X}\subseteq \mathbb{R}^n$ a set such that \eqref{asspt} holds. Let Assumption \ref{asspt:noise} hold.  
Consider 
the following SDP in the decision variables ${\color{black}P} \in \mathbb S^{n \times n}$,
$Y_1 \in \mathbb R^{T \times n}$, 
$G_2 \in \mathbb R^{T \times (s-n)}$, $\alpha \in \mathbb R_{> 0}${\color{black},} $\mu \in \mathbb R_{> 0}$:
\begin{subequations}
\label{eq:SDP-noise}
\begin{alignat}{6}
& {\color{black}P} \succ 0, \label{eq:SDP0-noise}\\
& Z_0 Y_1 = \begin{bmatrix} {\color{black}P} \\ 0_{(s-n) \times n} \end{bmatrix} \,,
\label{eq:SDP1-noise} \\
& 
\begin{bmatrix} 
M(Y_1, \alpha, \mu)& {\color{black}\star} & {\color{black}\star}  & {\color{black}\star}\\[1mm] 
{\color{black}G_2^\top X_1^\top}  & - I_{s-n} & {\color{black}\star}  & {\color{black}\star}\\
 R_Q^\top {\color{black}P} & 0_{{\color{black}\tau} \times n} & - I_{{\color{black}\tau}} & {\color{black}\star} \\[1mm] 
Y_1 & G_2 & 0_{T\times {\color{black}\tau}} & -\mu I_T
\end{bmatrix}  \preceq 0
\,, \label{eq:SDP2-noise} \\[1mm] 
& Z_0 G_2 = \begin{bmatrix} 0_{n \times (s-n)} \\ I_{s-n} \end{bmatrix}
\label{eq:SDP4-noise}
\end{alignat}
\end{subequations} 
where $M(Y_1, \alpha, \mu):=
X_1 Y_1 +(X_1 Y_1)^\top +
\alpha I_n + \mu E\Delta \Delta^\top E^\top$. If the program is feasible then the control law \eqref{state-feedback} with $K$ as in \eqref{eq:K_SDP}
is such that  the closed-loop dynamics \eqref{system-with-dist}, \eqref{state-feedback} are exponentially contractive on $\mathcal{X}$, i.e. \eqref{contractivity} holds. 
\end{theorem} 

\emph{Proof.} If  \eqref{eq:SDP-noise} is feasible, then we can define  $G_1:=Y_1{\color{black}P}^{-1}$,
and \eqref{eq:SDP1}, \eqref{eq:SDP4}  as well as  the definition of $K$ in  \eqref{eq:K_SDP}, imply that  \eqref{eq:GK} holds. Hence, by Lemma \ref{lem:main_noise},  the closed-loop dynamics $\dot x = (A+BK) Z(x)+Ed$ can be written as in \eqref{eq:GK_closed_noise}. By Schur complement, \eqref{eq:SDP2-noise} is equivalent to\footnote{To avoid burdening the notation, in the following expressions we omit the size of the zero matrices.}  
\[\ba{l}
\begin{bmatrix} X_1 Y_1 +(X_1 Y_1)^\top + \alpha I_n  & {\color{black}\star} & {\color{black}\star}\\[1mm] 
{\color{black}G_2^\top X_1^\top}  & - I_{s-n} & {\color{black}\star} \\
 R_Q^\top {\color{black}P} & 0 & - I_{{\color{black}\tau}}  \\[1mm] 
\end{bmatrix}  
\\
+
\mu 
\begin{bmatrix}
-E\\
0
\\
0
\end{bmatrix}
\Delta \Delta^\top
\begin{bmatrix}
-E\\
0
\\
0
\end{bmatrix}
^\top
+\mu^{-1}
\begin{bmatrix}
Y_1^\top\\
 G_2^\top\\
0
\end{bmatrix}
\begin{bmatrix}
Y_1^\top\\
 G_2^\top\\
0
\end{bmatrix}^\top
\preceq 0
\ea\]
which is compactly written as 
 $H+\mu L^\top \Delta \Delta^\top L+\mu^{-1} JJ^\top \preceq 0$, with obvious definition of $H,J, L$. Trivially, $L\ne 0$, whereas $\Delta \Delta^\top\succeq 0$. Moreover, $J\ne 0$, for otherwise $Y_1=0$, which would contradict \eqref{eq:SDP1-noise}.
Hence,  by the nonstrict Petersen's lemma \cite[Fact 2]{bisoffi2022data}, we 
have that $H+ J D^\top  L+
L^\top D J^\top\preceq 0$ for all $D\in \mathcal{D}$, i.e. 
\[\ba{l}
\begin{bmatrix} X_1 Y_1 +(X_1 Y_1)^\top + \alpha I_n  & {\color{black}\star} & {\color{black}\star}\\[1mm] 
{\color{black}G_2^\top X_1^\top}  & - I_{s-n} & {\color{black}\star} \\
 R_Q^\top {\color{black}P} & 0 & - I_{{\color{black}\tau}}  \\[1mm] 
\end{bmatrix}  
\\
+
{\color{black}\text{Sym}(\begin{bmatrix}
-E\\
0
\\
0
\end{bmatrix}
D
\begin{bmatrix}
Y_1 &
 G_2 &
0
\end{bmatrix})}
\preceq 0,\! \forall D\in \mathcal{D},
\ea\]
which can be rewritten as 
\[
\!\!\!\!
\ba{c}
\begin{bmatrix} 
{\color{black}\text{Sym}((X_1-ED) Y_1)} + \alpha I_n  
& {\color{black}\star} & {\color{black}\star}\\[1mm] 
G_2^\top (X_1-ED)^\top  & - I_{s-n} & {\color{black}\star} \\
 R_Q^\top {\color{black}P} & 0 & - I_{{\color{black}\tau}}  \\[1mm] 
\end{bmatrix}
\!\!  \preceq \! 0,\;
\forall D\in \mathcal{D}.
\ea\]
Observe now that this inequality is the same as \eqref{eq:SDP2} provided that $X_1 Y_1$ and $X_1 G_2$ in the latter are replaced by $(X_1-ED) Y_1$ and $(X_1-ED) G_2$. Hence, retracing the arguments of the proof of Theorem \ref{thm:contractivity}, one arrives at showing that 
\[\ba{l}
\!\!\!{\color{black}P}^{-1} (X_1-E D) G \frac{\partial Z(x)}{\partial x}+
\frac{\partial Z(x)}{\partial x}^\top G^\top (X_1-E D)^\top {\color{black}P}^{-1} \\
\hspace{2cm} \preceq -\beta {\color{black}P}^{-1}
\text{ for all $x\in \mathcal{X}$ and $D\in \mathcal{D}$}{\color{black}.}
\ea\]
Since $(X_1-E D_0) G \frac{\partial Z(x)}{\partial x}$ is the Jacobian of the closed-loop vector field in \eqref{eq:GK_closed_noise} and $D_0\in \mathcal{D}$ by Assumption \ref{asspt:noise}, the thesis is proven. \qedp

{\color{black}This result is the counterpart of Theorem \ref{thm:contractivity} in the case of perturbed data. Similar to Theorem \ref{thm:contractivity}, it gives a controller that guarantees  contractivity of the closed-loop system provided that an SDP is solved.  The use of perturbed data requires a more complex SDP, which depends on the size $\Delta$ of the perturbation. Moreover, its dimension is larger than the size of the noise-free SDP due to the  application of a matrix elimination lemma to get rid of the uncertainty introduced by the perturbation on the data. }

\smallskip

To state the asymptotic properties of the closed-loop system, we need to impose assumptions on the entire signal $d$ and not only on the samples collected during the experiment. Thus, we introduce the following:
\begin{assumption} \label{ass:RPI}
$d$ is a continuous function of time that,  for some known $\delta>0$ and for all $t\in \mathbb{R}$,  satisfies $|d(t)| \leq \delta$. \qedd
\end{assumption}  
A disturbance $d$ that satisfies Assumption \ref{ass:RPI} also satisfies Assumption \ref{asspt:noise} with $\Delta := \delta \sqrt{T} I_q$.

{\color{black}
\begin{rem}
It should be said that the choice of $\Delta$  above is only one possible choice and is made only for the sake of brevity. Another  choice  is based on using synthetic data obtained from pre-processing the data matrices $Z_0, X_1, U_0$. This pre-processing, which is detailed in \cite[Appendix ``Methods for data processing"]{dpt2023arc} and is inspired by \cite{BisoffiSCL2021tradeoffs},  leads to a bound $\Delta$ that is independent of the length $T$ of the dataset and  makes the size of the LMI independent of $T$ as well, which might have a positive impact on the feasibility of the LMI. Another option, also discussed in \cite[Appendix ``Methods for data processing"]{dpt2023arc}, 	consists of performing multiple experiments and then averaging the data. 
\end{rem}
}

The following result underscores that, by enforcing \eqref{contractivity},  one obtains that the closed-loop system is convergent (see Definition \ref{def:convergent-sys} in the Appendix) 
and as such it enjoys certain properties in the presence of forcing signals. 

 \begin{corollary}\label{cor2-noisy}
Let the conditions of Theorem \ref{thm:contractivity-noise} hold, with Assumption \ref{asspt:noise} replaced by Assumption \ref{ass:RPI} and $\Delta$ in \eqref{eq:SDP2-noise} replaced by $\delta \sqrt{T} I_q$, and where 
$\mathcal{X}=\mathbb{R}^n$. 
Then system $\dot x= (A+BK) Z(x)+E d$ is convergent. 
Let $x_*(t)$ be the unique solution that is defined and bounded for all ${\color{black}t\in \R}$ and globally uniformly exponentially stable. If $d(t)$ is periodic of period $T$ then $x_*(t)$ is periodic of period $T$. If $d(t)$ is constant, then $ x_*(t)=x_*$. {\color{black}Finally, for any $d(t)$ that satisfies Assumption \ref{ass:RPI} (hence, not necessarily periodic or constant), there exist a $\mathcal{K}\mathcal{L}$ function $\hat \beta$ and a $\mathcal{K}_\infty$ function $\hat \gamma$ such that the solution $x(t)$ of $\dot x = (A+BK)Z(x)+Ed+w$, where $w$ is a continuous and bounded input, satisfies 
\be\label{isc}\ba{l}
|x(t)-x_*(t)|\le \hat \beta(|x(0)-x_*(0)|,t) +\hat \gamma(\sup_{0\le \tau \le t}|w(t)|),\\
\hspace{3cm} \text{ for all $t\in \R_{\ge 0}$.}
\ea\ee
}
\end{corollary}

\smallskip

\emph{Proof.}  Under the stated conditions, the Jacobian of the right-hand side of  the closed-loop dynamics $\dot x= (A+BK) Z(x)+E d$, i.e. $(A+BK) \frac{\partial Z(x)}{\partial x}$,  satisfies \eqref{contractivity}. The {\color{black} first part of the} thesis is implied by Theorem \ref{th:demid-pavlov} in the Appendix. 
{\color{black}The final part of the thesis is given by \cite[Theorem 1]{pavlov2005convergent}.
\qedp
 

Corollary \ref{cor2-noisy} highlights that the controller designed in Theorem \ref{thm:contractivity-noise} endows the closed-loop system  with input-to-state stability, or input-to-state convergence according to the terminology in \cite{pavlov2005convergent}, with respect to the disturbances. 
If the disturbances are in addition periodic or constant, then the solutions of the closed-loop system designed from data converge to a unique solution, which is periodic or constant depending on the nature of the disturbance. In fact, if the disturbance is a linear combination of constant and sinusoidal signals of known frequencies, then Corollary \ref{cor2-noisy} holds with a simpler form of the SDP, as we show in {\color{black}Section \ref{sec:dist-known-freq}}. This feature will be exploited in the design of integral controllers in Section \ref{sec:integral-control}. 

{\color{black}
\begin{rem}
We are not aware of other approaches that deal with the data-driven {\em design} of controllers that enforce contractivity. As for the analysis, a so-called data-informativity approach has been proposed in \cite{ei2024cautious}, which tackles the problem of establishing  from data whether, for a given $P=P^\top \succ 0$ and $b\in \R$, a system  $\dot x= AZ(x)+d$ (with no control input) satisfies the condition
\be\label{int-demi-cond}
(Z(x)-Z(y))^\top A^\top P (x-y) \le b (x-y)^\top P (x-y),
\ee
for all $x,y\in \R^n$ (this is the integral version of the contractivity condition $(A\frac{\partial Z(x)}{\partial x})^\top P^{-1}+P^{-1}
A \frac{\partial Z(x)}{\partial x}\preceq b P^{-1}$). 
The samples of the dataset $\left\{ (x_i,\dot x_i) \right\}_{i=0}^{T-1}$ satisfy $X_1=AZ_0+D_0$, where $D_0\in \mathcal {D}$, and $\mathcal {D}$ models the class of disturbances. Under the conditions that $Z_0$ has full row rank and  the Lipschitz condition $\sqrt{(Z(x)-Z(y))^\top (Z_0Z_0^\top)^{-1} (Z(x)-Z(y))} \le L (x-y)^\top (x-y)$ holds for all $x,y\in \R^n$ and for some $L>0$, a condition is given (\cite[Corollary 4.2]{ei2024cautious}) that, if satisfied, proves \eqref{int-demi-cond}. 
The difference with respect to what is studied in this paper is not only formal, i.e. analysis vs.~design, but also substantial, because in our contribution, $P, K, \alpha$ {\em must} be designed, which is much harder. 
\end{rem}}

{\color{black}\subsection{Remainders}\label{sec:remainders}
We have shown above that contractivity of the closed-loop system can be still enforced when the plant is perturbed by an additive time-varying disturbance, provided that the convex program used to design the controller is suitably modified to take into account the perturbation through a bound. Other important  perturbations are the {\em state-dependent} ones. The study of these perturbations can be used to relax Assumption \ref{ass:Z}, which requires  the vector field $f(x)$ to belong to the finite-dimensional space of functions spanned by the elements of $Z(x)$. 

The main idea to relax Assumption 1 is to assume that $f(x)$ can be expressed as $AZ(x)+Ed(x)$, where $d\colon \R^n \to \R^n$ is a  continuously differentiable function and $E=I_n$. $d(x)$ models the nonlinearities that are not represented by the term $AZ(x)$. Hence, the term $d(x)$ can be regarded as a remainder or a neglected nonlinearity. As a result, the system is of the form
\begin{equation}\label{sys.with.remainder}
\dot x = A Z(x) + Bu +Ed(x). 
\ee
For this system, a dataset is collected as before and the samples due to $d(x)$ are stored in the matrix $D_0:=\begin{bmatrix}
d(x(t_0))&\ldots & d(x(x_{T-1}))\end{bmatrix}$. It is assumed that $D_0$ satisfies Assumption 3. Now, $d(x)$ is an extra nonlinearity to deal with. In agreement with  the approach pursued in the paper, we assume that the growth on the set $\mathcal{X}$ of the remainder $d(x)$ is known. Note that, we do not require $d(x)$ itself to be known, and hence we have to treat $d(x)$ differently from $Z(x)$ and do not use it in the feedback controller, whose structure remains unaltered, i.e. $u=KZ(x)$. Rather, we analyze how to preserve  contractivity of the closed-loop system in spite of the new term $d(x)$. 

Namely, we assume that there exists a matrix $R_D\in \R^{n\times \tau_d}$  such that 
\be\label{asspt-d(x)}
\frac{\partial d}{\partial x}(x)^\top \frac{\partial d}{\partial x}(x)\preceq R_D R_D^\top \text{ for all $x\in \mathcal{X}$}.
\ee
Under this condition, Theorem 2 can be adapted to deal with the presence of the remainder $d(x)$. In fact, the following holds:

\begin{theorem} \label{thm:contractivity-remainder}
Consider the nonlinear system  \eqref{sys.with.remainder} and let 
$Z(x)$ be of the form \eqref{eq:Z}. 
Let 
$ R_D\in \mathbb R^{n \times {\color{black}\tau}_d}$ be a known matrix and $\mathcal{X}\subseteq \mathbb{R}^n$ a set such that \eqref{asspt-d(x)} holds. Let Assumption \ref{asspt:noise} hold, where $D_0:=\begin{bmatrix}
d(x(t_0))&\ldots & d(x(x_{T-1}))\end{bmatrix}$.   
Consider 
the following SDP in the decision variables $P \in \mathbb S^{n \times n}$,
$Y_1 \in \mathbb R^{T \times n}$, 
$G_2 \in \mathbb R^{T \times (s-n)}$, $\alpha \in \mathbb R_{> 0}${\color{black},} $\mu \in \mathbb R_{> 0}$:
\begin{subequations}
\label{eq:SDP-remainder}
\begin{alignat}{6}
& \eqref{eq:SDP0-noise}, \eqref{eq:SDP1-noise}, \eqref{eq:SDP4-noise}\nonumber\\
& 
\begin{bmatrix} 
M(Y_1, \alpha, \mu)& \star & \star  & \star\\[1mm] 
\begin{bmatrix}G_2^\top X_1^\top  \\ I_n\end{bmatrix} & - I_{s} & \star  & \star\\
\begin{bmatrix} 
R_Q^\top \\[1mm]
R_D^\top
\end{bmatrix}  
 P & 0_{(\tau+\tau_d)\times s} & - I_{\tau+\tau_d} & \star \\[1mm] 
Y & \begin{bmatrix}G_2 & 0_{T\times n}\end{bmatrix} & 0_{T\times (\tau+\tau_d)} & -\mu I_T
\end{bmatrix}  \preceq 0
\,, \label{eq:SDP2-remainder} 
\end{alignat}
\end{subequations} 
where $M(Y_1, \alpha, \mu):=
X_1 Y_1 +(X_1 Y_1)^\top +
\alpha I_n + \mu \Delta \Delta^\top $. If the program is feasible then the control law \eqref{state-feedback} with $K$ as in \eqref{eq:K_SDP}
is such that  the closed-loop dynamics \eqref{sys.with.remainder}, \eqref{state-feedback}  are exponentially contractive on $\mathcal{X}$, i.e. 
{\color{black}
\be\label{contractivity-with-remainder}
\hspace{-3mm}
\begin{array}{l}
\exists P \in \mathbb S^{n \times n}, P \succ 0, \beta>0 \textrm{ such that } \forall x\in  \mathcal{X}\\
{\rm Sym}\left[
\left((A+BK)\displaystyle\frac{\partial Z(x)}{\partial x}+E \displaystyle\frac{\partial d(x)}{\partial x}\right)^\top P^{-1}
\right]  \preceq -\beta P^{-1}. 
\end{array}
\ee
}
\end{theorem} 

\emph{Proof.} 
{\color{black}
If  the above SDP is feasible, then we can define  $G_1:=Y_1P^{-1}$,
and by \eqref{eq:SDP1-noise}, \eqref{eq:SDP4-noise}  as well as  the expression of $K= U_0 \begin{bmatrix} Y_1 P^{-1} & G_2  \end{bmatrix}$,  the closed-loop dynamics $\dot x = (A+BK) Z(x)+d(x)$ can be written as
\begin{equation} \label{eq:GK_closed_reminder}
\dot x =  (X_1-D_0)G_1 x + [(X_1-D_0) G_2\; I_n] \begin{bmatrix}Q(x) \\ d(x)\end{bmatrix}.
\end{equation}
By Schur complement, \eqref{eq:SDP2-remainder} is equivalent to\footnote{To avoid burdening the notation, in the following expressions we omit the size of the zero matrices.}  
\[
\ba{l}
\begin{bmatrix} X_1 Y_1 +(X_1 Y_1)^\top + \alpha I_n  & 
\star 
& 
\star 
\\[1mm] 
\begin{bmatrix} G_2^\top X_1^\top  \\ I_n\end{bmatrix}  & - I_{s} & 
\star
\\
\begin{bmatrix} 
R_Q^\top \\[1mm]
R_D^\top
\end{bmatrix}  
 P & 0 & - I_{\tau+\tau_d}  \\[1mm] 
\end{bmatrix}  
\\
+
\mu 
\begin{bmatrix}
-I_n\\
0
\\
0
\\
0
\end{bmatrix}
\Delta \Delta^\top
\begin{bmatrix}
-I_n\\
0
\\
0
\\
0
\end{bmatrix}
^\top
+\mu^{-1}
\begin{bmatrix}
Y_1^\top\\
\begin{bmatrix}G_2^\top \\ 0_{n\times T}\end{bmatrix}\\
0
\end{bmatrix}
\begin{bmatrix}
Y_1^\top\\
 \begin{bmatrix}G_2^\top \\ 0_{n\times T}\end{bmatrix}\\
0
\end{bmatrix}^\top
\preceq 0
\ea
\]
which is compactly written as 
 $H+\mu L^\top \Delta \Delta^\top L+\mu^{-1} JJ^\top \preceq 0$, with obvious definition of $H,J, L$. Trivially, $L\ne 0$, whereas $\Delta \Delta^\top\succeq 0$. Moreover, $J\ne 0$, for otherwise $Y_1=0$, which would contradict \eqref{eq:SDP1-noise}.
Hence,  by the nonstrict Petersen's lemma \cite[Fact 2]{bisoffi2022data}, we 
have that $H+ J D^\top  L+
L^\top D J^\top\preceq 0$ for all $D\in \mathcal{D}$, i.e. 
\[
\ba{l} 
\begin{bmatrix} X_1 Y_1 +(X_1 Y_1)^\top + \alpha I_n  & 
\star 
& 
\star 
\\[1mm] 
\begin{bmatrix} G_2^\top X_1^\top  \\ I_n\end{bmatrix}  & - I_{s} & 
\star
\\
\begin{bmatrix} 
R_Q^\top \\[1mm]
R_D^\top
\end{bmatrix}  
 P & 0 & - I_{\tau+\tau_d}  \\[1mm] 
\end{bmatrix} \\  
+
\text{Sym}(\begin{bmatrix}
-I_n\\
0
\\
0
\\
0
\end{bmatrix}
D
\begin{bmatrix}
Y_1 &
  \begin{bmatrix}G_2 & 0_{T\times n}\end{bmatrix} &
0
\end{bmatrix})
\preceq 0, \quad \forall D\in \mathcal{D} 
\ea
\]
which can be rewritten as 
\[
\ba{l}
{\scriptsize\begin{bmatrix} (X_1-D) Y_1 + Y_1^\top (X_1-D)^\top + \alpha I_n  & 
\star
& 
\star 
\\[1mm] 
\begin{bmatrix}G_2^\top (X_1-D)^\top \\ I_n\end{bmatrix}  & - I_{s} &
\star
\\
\begin{bmatrix} 
R_Q^\top \\[1mm]
R_D^\top
\end{bmatrix}  
 P & 0 & - I_{\tau+\tau_d}  \\[1mm] 
\end{bmatrix}}
\!\!  \preceq \! 0,
\\ 
\hspace{3cm}   \forall D\in \mathcal{D}.
\ea
\]
Observe now that this inequality is the same as (12c) in the paper provided that $X_1 Y_1$, $X_1 G_2$ and $R_Q$ in the latter are replaced by $(X_1-D) Y_1$, $[(X_1-D) G_2\; I_n]$ and $[R_Q\; R_D]$ respectively. Hence, retracing the arguments of the proof of Theorem 1, one arrives at showing that 
\[
\ba{l}
\text{Sym}(P^{-1} ((X_1-D) G \frac{\partial Z(x)}{\partial x}+ \frac{\partial d(x)}{\partial x} ))
\preceq -\beta P^{-1} 
\\
\hspace{2cm} \text{ for all $x\in \mathcal{X}$ and $D\in \mathcal{D}$}.
\ea
\]
Since $(X_1-D_0) G \frac{\partial Z(x)}{\partial x} + \frac{\partial d(x)}{\partial x}$ is the Jacobian of the closed-loop vector field in \eqref{eq:GK_closed_reminder} and $D_0\in \mathcal{D}$ by Assumption \ref{asspt:noise}, the thesis is proven.}

\qedp

The result shows that one can enforce contractivity by feedback even without  Assumption \ref{ass:Z},  provided that 
the remainder $d(x)$ that measures the discrepancy $f(x)-A Z(x)$ 
 has a bounded Jacobian on the set $\mathcal{X}$ of interest. 
{\color{black} As the closed-loop system $\dot x = (A +BK)Z(x)+Ed(x)$ is contractive, it enjoys the asymptotic properties of any contractive system. Namely, by Theorem \ref{th:demid-pavlov}, if  $\mathcal{X}=\R^n$ then there exists a unique  equilibrium $x_*$, which is globally exponentially stable. 
Moreover, by Corollary \ref{cor:convergent}, if  $\mathcal{X}$ is a convex subset of $\R^n$ and there exists $\overline x\in \text{int}(\mathcal{X})$ such that $(A +BK)Z(\overline x)+Ed(\overline x)=0$, then $\overline x$ is  exponentially stable. }  

\begin{rem}
{\color{black}
Systems of the form \eqref{sys.with.remainder} naturally arise from 
estimation techniques, notably 
kernel-based estimation (see, e.g., \cite{maddalena2021deterministic} and 
\cite{hu2023learning}). Consider a positive semidefinite kernel $K\colon \mathcal{X}\times\mathcal{X}\to \R$, 
and the associated RKHS $\mathcal{H}$. Assume that all the $n$ entries of the vector field $f$ belong to $\mathcal{H}$.  Using the dataset $\left\{ (x_i,\dot x_i) \right\}_{i=0}^{T-1}$, where the samples satisfy $\dot x_i = f(x_i)$ for each $i=0,1,\ldots, T-1$, 
one 
determines the interpolation function $A Z(x)$ for $f(x)$
such that $f(x)= A Z(x) +d(x)$, where each entry $d_i(x)$ of the remainder $d(x)$ satisfies
${\color{black}|d_i(x)|  \leq} \| f_i(x)\|_{\mathcal H} \kappa(x)$,
and $\kappa(x)$ is a known function that depends on the kernel $K$  and the dataset. The factor $\| f_i(x)\|_{\mathcal H}$ is unknown but it is usually approximated by an upper bound. 
The main obstacle that prevents the use of Theorem \ref{thm:contractivity-remainder} on the  system $A Z(x) +B u+d(x)$ is that it requires an upper bound on the Jacobian of $d(x)$ while the kernel-based estimate returns a bound on $d(x)$ itself. Enforcing contraction while using kernel-based estimates is left for future research. }
\end{rem}
 
}
\section{Data generated under disturbances of known frequencies}\label{sec:dist-known-freq}

In this subsection, we investigate the case in which the disturbance perturbing the system is a linear combination of constant and sinusoidal signals of known frequencies. 
{\color{black}This class of signals is relevant in many engineering problems, including output regulation \cite{byrnes1997output} and vibration control \cite{landau2016adaptive}. In the next section, we show how the result established here can be used in the data-based design of integral controllers for nonlinear systems.}

Formally, we replace Assumption \ref{ass:RPI} with the following one:
\begin{assumption} \label{ass:periodic-dist-exosys}
$d$ is a function of time solution of the  autonomous system
\be\label{exosystem}
\dot w = \Psi w, \quad d= \Gamma w{\color{black},}
\ee
where {\color{black}$w=\begin{bmatrix}w_1^\top\, \ldots w_{\sigma_1}^\top\,  w_{\sigma_1+1}\, \ldots, w_{\sigma_1+\sigma_2}\end{bmatrix}^\top\in \R^q$}, 
\[
\Psi=
{\color{black}{\rm diag}}
(\left[\begin{smallmatrix} 0 & \psi_1\\ -\psi_1 & 0 \end{smallmatrix}\right],\ldots,
\left[\begin{smallmatrix} 0 & \psi_{\sigma_1}\\ -\psi_{\sigma_1} & 0 \end{smallmatrix}\right], \underbrace{0,\ldots, 0}_{\text{$\sigma_2$ times}}),
\]
$\psi_1, \ldots, \psi_{\sigma_1}\in \mathbb{R}_{>0}$ known frequencies, and $\Gamma\in \mathbb{R}^{q\times (2\sigma_1+\sigma_2)}$  an unknown matrix. 
\qedd
\end{assumption}
{\color{black} In the Assumption above, the integer $\sigma_1$ is the number of pairs of complex conjugate eigenvalues of the matrix $\Psi$ and the integer $\sigma_2$ is the number of its real eigenvalues. Accordingly, $w_i:=[w_{i1}\, w_{i2}\,]^\top$  for $i=1,2,\ldots, \sigma_1$ are $2$-dimensional sub-vectors of $w$, while $w_i$  for $i=\sigma_1+1,\sigma_1+2,\ldots, \sigma_1+\sigma_2$ are scalar entries of $w$.}

We denote by 
\[
w(0)=\begin{bmatrix}w_1(0)^\top\!\!\!\!&\ldots&\!\!\!\!
w_{\sigma_1}(0)^\top&\!\!\!\! w_{\sigma_1+1}(0)&\!\!\ldots& 
\!\!w_{\sigma_1+\sigma_2}(0)\end{bmatrix}^\top
\] 
the initial condition of the system \eqref{exosystem} generating the disturbance during the data collection phase, where $w_i(0)=[w_{i1}(0) \; w_{i2}(0)]^\top\in \mathbb{R}^2$ for $i=1, \ldots, \sigma_1$ and 
$w_i(0)\in \mathbb{R}$ for $i=\sigma_1+1, \ldots, \sigma_1+\sigma_2$. Under Assumption \ref{ass:periodic-dist-exosys} the matrix of disturbance samples $D_0$ in \eqref{eq:data5} becomes
\[
D_0 = \Gamma
\begin{bmatrix}
w_1(t_0) & \ldots & w_1(t_{T-1}) \\
\vdots & \ddots & \vdots\\
w_{\sigma_1+\sigma_2}(t_0) & \ldots & w_{\sigma_1+\sigma_2}(t_{T-1}) \\
\end{bmatrix}
\]
where, for $i=1, 2,\ldots, \sigma_1$, 
\[\ba{rl}
& \begin{bmatrix}
w_i(t_0) & \ldots & w_i(t_{T-1}) 
\end{bmatrix}\\[3mm]
=&
\underbrace{
\begin{bmatrix}
w_{i1}(0) & w_{i2}(0)\\
w_{i2}(0) & -w_{i1}(0)
\end{bmatrix}
}_{=:L_i}
\underbrace{
\begin{bmatrix}
\cos(\psi_i t_0) & \ldots & \cos(\psi_i t_{T-1})\\
\sin(\psi_i t_0) & \ldots & \sin(\psi_i t_{T-1})\\ 
\end{bmatrix}
}_{=:{\color{black}W}_i}
\ea
\]
and, for $i=\sigma_1+1,\sigma_1+ 2,\ldots, \sigma_1+\sigma_2$,
\[
\begin{bmatrix}
w_i(t_0) & \ldots & w_i(t_{T-1}) 
\end{bmatrix}
=
\underbrace{
w_{i}(0)
}_{=:L_i}
\underbrace{
\begin{bmatrix}
1 & \ldots & 1
\end{bmatrix}
}_{=:{\color{black}W}_i}.
\]
Set 
\be\label{M}
\ba{rl}
L:=&
{\color{black}{\rm diag}}(L_1, \ldots, L_{\sigma_1+\sigma_2})
{\color{black}\in \R^{(2\sigma_1+\sigma_2)\times (2\sigma_1+\sigma_2)}}
\\[1mm]
{\color{black}W}:=&
{\color{black}
\begin{bmatrix}
{\color{black}W}_1^\top\, \ldots \, {\color{black}W}_{\sigma_1+\sigma_2}^\top
\end{bmatrix}^\top \in \R^{(2\sigma_1+\sigma_2)\times T}
}
\\[1mm]
{\color{black}\Phi}:=&\Gamma L{\color{black}\in \R^{q\times (2\sigma_1+\sigma_2)}}
\ea
\ee
and note that ${\color{black}W}$ is known while ${\color{black}\Phi}$ is unknown. Bearing in mind the system \eqref{system2-with-dist}, and the expression of $D_0$ given above, namely $D_0=\Gamma L {\color{black}W}={\color{black}\Phi W}$, we realize that the matrices of data satisfy 
\[
X_1= A Z_0+B U_0+ E {\color{black}\Phi W}. 
\]
Hence, we have the following interesting data-dependent representation of the  closed-loop system 
\begin{lem} \label{lem:main_noise_exosys}
Let Assumption \ref{ass:Z} and \ref{ass:periodic-dist-exosys}
hold.
Consider any matrices $K \in \mathbb R^{m \times s}$,
$G \in \mathbb R^{T \times s}$ such that 
\begin{equation} \label{eq:GK-exosystem}
\begin{bmatrix} K \\ I_s \\ 0_{(2 \sigma_1+\sigma_2) \times s}\end{bmatrix} = \begin{bmatrix} U_0 \\ Z_0 \\ {\color{black}W}\end{bmatrix} G
\end{equation}
 holds. 
Let $G$ be partitioned as $G = \begin{bmatrix} G_1 & G_2 \end{bmatrix}$, where
$G_1 \in \mathbb R^{T \times n}$ and $G_2 \in \mathbb R^{T \times (s-n)}$.
Then system \eqref{system-with-dist} under the control law 
\eqref{state-feedback}
results in
the closed-loop dynamics 
\begin{equation} \label{eq:GK_closed_noise_periodic}
\!\!\!\!\!\!
\ba{rl}
\dot x =& (A+BK) Z(x) +Ed \\
=& X_1 G_1 x + X_1G_2  Q(x) +Ed. 
\ea\end{equation}
\end{lem} 

\emph{Proof.} Left-multiply both sides of \eqref{eq:GK-exosystem} by $\left[\begin{smallmatrix}B & A & E{\color{black}\Phi} \end{smallmatrix}\right]$ and obtain
\[
A+BK = (A Z_0 + BU_0 +E{\color{black}\Phi W} )G = X_1 G,
\]
{\color{black} where the second identity holds thanks to the condition $0_{(2 \sigma_1+\sigma_2) \times s}={\color{black}W}G$, which guarantees that $E{\color{black}\Phi W}G=0_{n\times s}$.}
This ends the proof. \qedp

\smallskip

The remarkable feature of this result is that, for systems affected by disturbances generated by \eqref{exosystem}, 
a data-dependent representation that is {\em independent} of the unknown matrix $D_0$ is derived, provided that the additional condition $0_{{\color{black}(2\sigma_1+\sigma_2)}\times s}= {\color{black}W} G$ holds. This condition can be more or less restrictive depending on $W$, whose row dimension increases with the number of frequencies to be rejected. As a result, rejecting a constant disturbance (a bias) will generally be easier than rejecting a signal containing multiple frequencies. The result allows us to derive contractivity and asymptotic properties of the closed-loop system {\color{black} under conditions} similar to those 
{\color{black} considered} 
in the case of noise-free data, {\color{black} making their fulfilment easier than in the case of data perturbed by a ``general" noise  considered in Section \ref{sec:noisy-data}}. These results are introduced below. 

\begin{theorem} \label{thm:contractivity-periodic-noise}
Consider the nonlinear system  \eqref{system-with-dist}, let Assumption \ref{ass:Z} hold 
and $Z(x)$ be of the form \eqref{eq:Z}. Let 
$ R_Q\in \mathbb R^{n \times \tau}$ be a known matrix and $\mathcal{X}\subseteq \mathbb{R}^n$ a set such that \eqref{asspt} holds.
Let Assumption \ref{ass:periodic-dist-exosys} hold.  
Consider 
the following SDP in the decision variables ${\color{black}P} \in \mathbb S^{n \times n}$,
$Y_1 \in \mathbb R^{T \times n}$, 
$G_2 \in \mathbb R^{T \times (s-n)}$, $\alpha \in \mathbb R_{> 0}$:
\begin{equation}
\label{eq:SDP-noise-exosys}
{\rm \eqref{eq:SDP}},\; 0_{(2 \sigma_1+\sigma_2)\times s}= {\color{black}W}\begin{bmatrix} Y_1 & G_2 \end{bmatrix},
\end{equation} 
where ${\color{black}W}$ is the matrix defined in \eqref{M}.
If the program is feasible then the control law \eqref{state-feedback} with $K$ as in \eqref{eq:K_SDP}
is such that  the closed-loop dynamics \eqref{system-with-dist}, \eqref{state-feedback} are exponentially contractive on $\mathcal{X}$, i.e. \eqref{contractivity} holds. 
\end{theorem} 

\emph{Proof.} If  \eqref{eq:SDP-noise-exosys} is feasible then \eqref{eq:SDP} is feasible. Then  we can define  $G_1:=Y_1{\color{black}P}^{-1}$
and note that \eqref{eq:SDP1}, \eqref{eq:SDP4},  the definition of $K$ in  \eqref{eq:K_SDP} and the constraint $0_{(2 \sigma_1+\sigma_2)\times s}= {\color{black}W}\begin{bmatrix} Y_1 & G_2 \end{bmatrix}$ give \eqref{eq:GK-exosystem}. Hence, by Lemma \ref{lem:main_noise_exosys},  the closed-loop dynamics $\dot x = (A+BK) Z(x)+Ed$ can be written as $\dot x = Mx + N Q(x)+Ed$, with 
$M=X_1G_1$ and $N=X_1G_2$. 
By \eqref{eq:SDP2} and \eqref{asspt}, the proof of Theorem \ref{thm:contractivity} shows that \eqref{contractivity} holds. 
\qedp

{\color{black}Under the conditions of Theorem \ref{thm:contractivity-periodic-noise}, the closed-loop system is exponentially contractive, hence its solutions entrain to periodic solutions when $d(t)$ is periodic, similar to Corollary \ref{cor2-noisy}. Note, however,  that entrainment is obtained under conditions that are less demanding  than those in Corollary \ref{cor2-noisy}.
}

%
%

\section{Data driven integral control for nonlinear systems}\label{sec:integral-control}

In this section we discuss  how establishing contractivity via data paves the way for the data-driven design of an integral controller for nonlinear systems of the form
\begin{equation}
\label{system-with-disturbance}
\ba{rl}
\dot x =& f(x) + B u +Ed\\
y= & h(x)\\
e=& y-r
\ea
\end{equation}  
where $d\in \mathbb{R}^q,r\in \mathbb{R}^p$ are a {\em constant} disturbance and a  {\em constant}  reference signal, $y\in \mathbb{R}^p$ is the regulated output and $e\in \mathbb{R}^p$ is the tracking error. We assume that $y$ is also available for measurements, in addition to the state $x$. 
We are interested in the problem of designing the integral controller 
\be\label{integral-controller}\ba{rl}
\dot \xi = & e\\
u = & k(x,\xi)
\ea
\ee
{\color{black} where $\xi\in \R^p$ is the state of the integral controller,}\footnote{{\color{black} The variable $\xi$  here should not be confused with the one in Section IV.}}
such that the solutions $x(t), \xi(t)$ of the closed-loop system 
\be\label{closed-loop-integral-controller}\ba{rl}
\dot x =& f(x) + B k(x,\xi)+Ed\\
\dot \xi = & e\\
e = & h(x)-r
\ea
\ee
are bounded for all $t\ge 0$ and satisfy $e(t)\to 0$ as $t\to +\infty$. 

{\color{black} Note that integral controllers are common in control applications, and having a method to ``tune'' them from data has always been a primary concern. In this section we provide a ``tuning'' method for the nonlinear system \eqref{system-with-disturbance}. }

The following assumption on the system replaces Assumption \ref{ass:Z}.

\begin{assumption} \label{ass:Z-with-output}
A continuously differentiable vector-valued function $Z: \mathbb R^n \rightarrow \mathbb R^s$ is known such that 
$f(x) = AZ(x)$ and $h(x)=CZ(x)$ for some matrices $A \in \mathbb R^{n \times s}$, $C \in \mathbb R^{p \times s}$. \qedd
\end{assumption}

Hence, the system \eqref{system-with-disturbance} considered in this section takes the form 
\begin{equation}\label{sys-w-output}
\ba{rl}
\dot x = & A Z(x) + B u+Ed\\
e= & CZ(x)-r.
\ea
\end{equation}  
We let $Z(x)$ be of the form \eqref{eq:Z} and partition $A,C$ accordingly as
\[
\begin{bmatrix}
A \\
C
\end{bmatrix}
=
\begin{bmatrix}
\,\overline A &  \hat A\\
\,\overline C & \hat C
\end{bmatrix}.
\]
After stacking  $Z(x)$ and the controller state variable $\xi$ in the vector 
\be\label{calZ}
\mathcal{Z}(x, \xi)=
\begin{bmatrix}
x\\
\xi\\
Q(x)
\end{bmatrix},
\ee
the system \eqref{system-with-disturbance}  extended with the integral action $\dot \xi= e$ can be written as 
\be\label{extended-system}
\ba{rl}
\begin{bmatrix}
\dot x\\
\dot \xi
\end{bmatrix}
=& \mathcal{A} \mathcal{Z}(x,\xi)+\mathcal{B} u 
+\mathcal{E} d -\mathcal{I} r
\ea
\ee
where
\[\ba{c}
\mathcal{A}=\begin{bmatrix}
\,\overline A & 0_{n\times p} & \hat A\\
\,\overline C & 0_{p\times p} & \hat C
\end{bmatrix}, \; 
\mathcal{B}=\begin{bmatrix}
B \\
0_{p\times m}
\end{bmatrix}\\
\mathcal{E}=\begin{bmatrix}
E \\
0_{p\times q}
\end{bmatrix}
,\; \mathcal{I} =\begin{bmatrix}
0_{n\times p} \\
I_p
\end{bmatrix}.
\ea\]
The constant disturbance $d$ is generated by the system \eqref{exosystem} with $\Sigma=0_{\sigma\times \sigma}$, and $\Gamma \in \mathbb{R}^{q\times \sigma}$ unknown.

We collect data 
\begin{equation} \label{dataset-extended}
\mathbb D := \left\{ (x_i,u_i, y_i, \dot x_i, \xi_i)\right\}_{i=0}^{T-1}
\end{equation} 
from systems \eqref{system-with-disturbance}, \eqref{integral-controller} through an experiment, where, as before,  $x_i:= x(t_i)$, etc., 
 and define, in addition to the data matrix $U_0$ defined in \eqref{eq:data}, 
the matrices
\[\ba{rl}
\mathcal{Z}_1:=& \begin{bmatrix}
\dot x(t_0) & \dot x(t_1) & \ldots & \dot x(t_{T-1})\\ 
y(t_0) & y(t_1) & \ldots & y(t_{T-1})\\ 
\end{bmatrix}\\[4mm]
\mathcal{Z}_0:=& \begin{bmatrix}
x(t_0) &  x(t_1) & \ldots &  x(t_{T-1})\\ 
\xi(t_0)&  \xi(t_1) & \ldots & \xi(t_{T-1})\\ 
Q(x(t_0)) &  Q(x(t_1)) & \ldots &  Q(x(t_{T-1}))\\ 
\end{bmatrix}.
\ea\]
Bearing in mind the analysis of Section \ref{sec:dist-known-freq}, the matrix of disturbance samples $D_0$ in \eqref{eq:data5} is equal to ${\color{black}\Phi W}$ where ${\color{black}\Phi}=\Gamma L$,  
\be\label{L-constant}
L={\rm diag}(w_{1}(0), \ldots, w_{\sigma}(0))\in \mathbb{R}^{\sigma\times \sigma}
\ee
\be\label{M-constant}
{\color{black}W}=
\left[\begin{smallmatrix}
\mathds{1}_{1\times T}\\ \vdots \\ \mathds{1}_{1\times T}
\end{smallmatrix}\right]
\in \mathbb{R}^{\sigma\times T}
\ee
and $\mathds{1}_{1\times T}$ denotes a $1\times T$ matrix of all ones. 
The matrices of data satisfy the identity 
\[
\mathcal{Z}_1= 
\mathcal{A}\mathcal{Z}_0
+
\mathcal{B}U_0
+
\mathcal{E}{\color{black}\Phi W}.
\]
Note that to derive this identity, we are using the relation
\[
y(t_i)=e(t_i)+r=C Z(x(t_i)),\; i=0,1,\ldots, T-1.
\]

We can give a data-dependent representation of system \eqref{extended-system} in feedback with 
\be\label{K-integral-controller}\ba{rl}
u=k(x,\xi):=
\begin{bmatrix}
\,\overline K & \widetilde K & \hat K
\end{bmatrix}
\begin{bmatrix}
x\\
\xi\\
Q(x)
\end{bmatrix}
=:
\mathcal{K} \mathcal{Z}(x,\xi).
\ea\ee

\begin{lem} \label{lem:integral_control}
Consider any matrices $\mathcal{K} \in \mathbb R^{m \times (s+p)}$,
$\mathcal{G} \in \mathbb R^{T \times (s+p)}$ such that \begin{equation} \label{eq:GcalK}
\begin{bmatrix} \mathcal{K}  \\ I_{s+p}\\ 0_{\sigma \times (s+p)} \end{bmatrix} = \begin{bmatrix} U_0 \\ \mathcal{Z}_0\\ {\color{black} W} \end{bmatrix} \mathcal{G}
\end{equation}
 holds. 
Let $\mathcal{G}$ be partitioned as $\mathcal{G} = \begin{bmatrix} \mathcal{G}_1 & \mathcal{G}_2 \end{bmatrix}$, where
$\mathcal{G}_1 \in \mathbb R^{T \times (n+p)}$ and $\mathcal{G}_2 \in \mathbb R^{T \times (s-n)}$.
Then, the dynamics of the system \eqref{system-with-disturbance}, \eqref{integral-controller}, where $k(x, \xi)$ is as in \eqref{K-integral-controller},
result in
the closed-loop dynamics 
%
\be\label{extended-system-closed}
\ba{l}
\begin{bmatrix}
\dot x\\
\dot \xi
\end{bmatrix}
= (\mathcal{A} +\mathcal{B} \mathcal{K}) \mathcal{Z}(x,\xi)
+\mathcal{E} d -\mathcal{I} r
\\
= 
\mathcal{Z}_1\mathcal{G}_1 \begin{bmatrix}
x\\
\xi
\end{bmatrix}
+ \mathcal{Z}_1\mathcal{G}_2  Q(x) +\mathcal{E} d -\mathcal{I} r.
\ea
\ee
\end{lem}

\emph{Proof.} The proof is similar to the proof of Lemma \ref{lem:main_noise_exosys} and is omitted.  
\qedp

The result below shows contractivity of the closed-loop dynamics.

\begin{theorem} \label{thm:contractivity-extended}
Consider the nonlinear system  \eqref{system-with-disturbance}, let Assumption \ref{ass:Z-with-output} hold 
and $\mathcal{Z}(x,\xi)$ be of the form \eqref{calZ}. 
Let 
$ R_Q\in \mathbb R^{n \times \tau}$ be a known matrix and $\mathcal{X}\subseteq \mathbb{R}^n$ a set such that \eqref{asspt} holds.
Set $\mathcal{R}_Q:=\left[\begin{smallmatrix} R_Q \\ 0_{p\times {\color{black}\tau}}\end{smallmatrix}\right]$.
Consider 
the following SDP in the decision variables ${\color{black}\mathcal{P}} \in \mathbb S^{(n+p) \times (n+p)}$,
$\mathcal{Y}_1 \in \mathbb R^{T \times (n+p)}$, 
$\mathcal{G}_2 \in \mathbb R^{T \times (s-n)}$, $\alpha \in \mathbb R_{> 0}$: 
\begin{subequations}
\label{eq:SDP-extended}
\begin{alignat}{6}
& {\color{black}\mathcal{P}} \succ 0, \label{eq:SDP0-extended}\\
& \mathcal{Z}_0 \mathcal{Y}_1 = \begin{bmatrix} {\color{black}\mathcal{P}} \\ 0_{(s-n) \times (n+p)} \end{bmatrix} \,,
\label{eq:SDP1-extended} \\
& 
\begin{bmatrix} 
\mathcal{Z}_1 \mathcal{Y}_1 +(\mathcal{Z}_1 \mathcal{Y}_1)^\top + 
\alpha I_{n+p}
& {\color{black}\star} & {\color{black}\star}  \\[1mm] 
{\color{black}\mathcal{G}_2^\top \mathcal{Z}_1^\top} & - I_{s-n} & {\color{black}\star}  \\
 \mathcal{R}_Q^\top {\color{black}\mathcal{P}} & 0 & - I_{{\color{black}\tau}} 
\end{bmatrix}  \preceq 0
\,, \label{eq:SDP2-extended} \\[1mm] 
& \mathcal{Z}_0 \mathcal{G}_2 = \begin{bmatrix} 0_{(n+p) \times (s-n)} \\ I_{s-n} \end{bmatrix},
\label{eq:SDP4-extended}\\
& \mathds{1}_{1\times T} \begin{bmatrix} \mathcal{Y}_1 & \mathcal{G}_2 \end{bmatrix}  = 0_{1 \times (s+p)}  \,. \label{eq:SDP5-extended} 
\end{alignat}
\end{subequations} 
If the program is feasible then the control law \eqref{K-integral-controller} 
with $\mathcal{K}$ defined as 
\begin{eqnarray} \label{eq:K_SDP-extended}
\mathcal{K}= U_0 \begin{bmatrix} \mathcal{Y}_1 {\color{black}\mathcal{P}}^{-1}
& \mathcal{G}_2 \end{bmatrix}
\end{eqnarray}
is such that  the closed-loop dynamics \eqref{system-with-disturbance}, \eqref{integral-controller}  are exponentially contractive on $\mathcal{X}\times \mathbb{R}^p$, i.e. 
\[
\!\!\!
\ba{l}
\exists \beta>0 \text{ such that for all } (x,\xi)\in \mathcal{X}\times \mathbb{R}^p\\
\left[(\mathcal{A} +\mathcal{B} \mathcal{K}) \frac{\partial \mathcal{Z}{\color{black}(x,\xi)}}{\partial (x,\xi)}\right]^\top \!\!\!
{\color{black}\mathcal{P}}^{-1}
+{\color{black}\mathcal{P}}^{-1}\!\!
\left[(\mathcal{A} +\mathcal{B} \mathcal{K}) \frac{\partial \mathcal{Z}{\color{black}(x,\xi)}}{\partial (x,\xi)}\right]\!\!\preceq \!-\beta {\color{black}\mathcal{P}}^{-1}.
\ea\] 
\end{theorem}

{\emph Proof.} 
If \eqref{eq:SDP-extended} is feasible and we set $\mathcal{G}_1=\mathcal{Y}_1{\color{black}\mathcal{P}}^{-1}$, then it holds that 
\be\label{int-ineq-extended-0}
\ba{l}
{\color{black}\mathcal{P}}^{-1} \mathcal{Z}_1  \mathcal{G}_1 +  \mathcal{G}_1^\top  \mathcal{Z}_1^\top{\color{black}\mathcal{P}}^{-1}  + \alpha {\color{black}\mathcal{P}}^{-2} +  \mathcal{R}_Q  \mathcal{R}_Q^\top \\[1mm]
 \hspace{0.5cm} + {\color{black}\mathcal{P}}^{-1} \mathcal{Z}_1 \mathcal{G}_2 
 \mathcal{G}_2^\top \mathcal{Z}_1   {\color{black}\mathcal{P}}^{-1}  \preceq 0.
\ea
\ee
If $\mathcal{Z}_1 \mathcal{G}_2=0$, then $(\mathcal{A} +\mathcal{B} \mathcal{K}) \frac{\partial \mathcal{Z}(x,\xi)}{\partial (x,\xi)}= \mathcal{Z}_1\mathcal{G}_1 \left[\begin{smallmatrix} x\\ \xi \end{smallmatrix}\right]$, and \eqref{int-ineq-extended-0} implies the exponential contractivity property with $\beta:=\alpha \lambda_{\min}({\color{black}\mathcal{P}}^{-1})$, as claimed. 
If $\mathcal{Z}_1 \mathcal{G}_2\ne 0$, \eqref{int-ineq-extended-0} implies
that 
\be\label{int-ineq-extended-implied-by}
\ba{l}
{\color{black}\text{Sym}(\mathcal{P}^{-1} \mathcal{Z}_1\mathcal{G}_1 +\mathcal{P}^{-1} \mathcal{Z}_1\mathcal{G}_2 \mathcal{R}^\top)  + \alpha \mathcal{P}^{-2}}\preceq 0,\\[2mm]
\text{ for all }
\mathcal{R}\in \{\mathcal{R}\colon \mathcal{R}\mathcal{R}^\top\preceq \mathcal{R}_Q \mathcal{R}_Q^\top\}.
\ea\ee
By \eqref{asspt}, for any $(x,\xi)\in \mathcal{X}\times \mathbb{R}^p$, we have
\[\ba{l}
\displaystyle
\frac{\partial Q{\color{black}(x,\xi)}}{\partial (x,\xi)}^\top
\frac{\partial Q{\color{black}(x,\xi)}}{\partial (x,\xi)}=
\begin{bmatrix}
\frac{\partial Q{\color{black}(x,\xi)}}{\partial x}^\top
\\
\frac{\partial Q{\color{black}(x,\xi)}}{\partial \xi}^\top
\end{bmatrix}
\begin{bmatrix}
\frac{\partial Q{\color{black}(x,\xi)}}{\partial x}&
\frac{\partial Q{\color{black}(x,\xi)}}{\partial \xi}
\end{bmatrix}
\preceq \\[5mm]
\hspace{3.5cm}
\begin{bmatrix}
R_Q R_Q^\top& 0_{}\\
0_{} & 0_{}
\end{bmatrix}
=\mathcal{R}_Q \mathcal{R}_Q^\top
\ea\]
that is, $\frac{\partial Q{\color{black}(x,\xi)}}{\partial (x,\xi)}^\top \in \{\mathcal{R}\colon \mathcal{R}\mathcal{R}^\top\preceq \mathcal{R}_Q \mathcal{R}_Q^\top\}$. Therefore 
\[
\ba{l}
{\color{black}\text{Sym}(\mathcal{P}^{-1} \mathcal{Z}_1\mathcal{G}_1 +\mathcal{P}^{-1} \mathcal{Z}_1\mathcal{G}_2\frac{\partial Q{\color{black}(x,\xi)}}{\partial (x,\xi)})  + \alpha \mathcal{P}^{-2}}\preceq 0
\ea
\]
that is, recalling that $\beta=\alpha \lambda_{\min}({\color{black}\mathcal{P}}^{-1})$, 
\[\ba{l}
{\color{black}\mathcal{P}}^{-1} \mathcal{Z}_1 \mathcal{G} \frac{\partial \mathcal{Z}{\color{black}(x,\xi)}}{\partial (x,\xi)}
+
\frac{\partial \mathcal{Z}{\color{black}(x,\xi)}}{\partial (x,\xi)}^\top \mathcal{G}^\top  \mathcal{Z}_1^\top {\color{black}\mathcal{P}}^{-1}\preceq -\beta {\color{black}\mathcal{P}}^{-1}
\ea\]
for all $(x,\xi)\in \mathcal{X}\times \mathbb{R}^p$. Bearing in mind that, by \eqref{eq:SDP0-extended}, \eqref{eq:SDP1-extended}, \eqref{eq:SDP4-extended}, \eqref{eq:SDP5-extended}, \eqref{eq:K_SDP-extended} and Lemma \ref{lem:integral_control}, 
\[
(\mathcal{A} +\mathcal{B} \mathcal{K}) \mathcal{Z}(x,\xi)=
\mathcal{Z}_1\mathcal{G}_1 \begin{bmatrix}
x\\
\xi
\end{bmatrix}
+ \mathcal{Z}_1\mathcal{G}_2  Q(x),
\]
the thesis is complete.  \qedp

The application to the data-driven integral control problem is immediate.
\begin{corollary}\label{cor:integr-control}
Let the conditions of Theorem \ref{thm:contractivity-extended} with $\mathcal{X}=\mathbb{R}^n$ hold.
Then the control law \eqref{K-integral-controller} 
with $\mathcal{K}$ defined as in \eqref{eq:K_SDP-extended}
is such that  the solutions of the closed-loop dynamics \eqref{system-with-disturbance}, \eqref{integral-controller} are bounded for all $t\ge 0$ and satisfy $e(t)\to 0$ as $t\to +\infty$. 
\end{corollary} 

{\em Proof.} By Theorem \ref{thm:contractivity-extended}, the closed-loop dynamics \eqref{system-with-disturbance}, \eqref{integral-controller}, where $k(x,\xi)$ is as in 
\eqref{K-integral-controller} and $\mathcal{K}$ is as in  \eqref{eq:K_SDP-extended}, is exponentially contractive on $\mathbb{R}^n\times\mathbb{R}^p$. Moreover, it is time invariant. By Theorem \ref{th:demid-pavlov} in the Appendix, all its solutions are defined for all $t\in \mathbb{R}_{\ge 0}$ and for all $(x(0), \xi(0))\in \mathbb{R}^n\times\mathbb{R}^p$. There exists a unique equilibrium  $(x_\star, \xi_\star)$, which is globally uniformly exponentially stable. Hence, all the solutions are bounded. Finally, since $\dot \xi = e$, at the equilibrium $0 = h(x_*)-r$, hence by continuity of $h$, $e(t)\to 0$ as $t\to +\infty$.  \qedp

\smallskip

The result just stated gives conditions for the regulation of the output of a nonlinear system to a constant reference, {\color{black}while simultaneously rejecting constant disturbances}. Given the nature of the result, one would expect some conditions on the feasibility on the problem to be present in the statement. For instance, given that the constant reference value $r$ can be any vector in $\mathbb{R}^p$, it would be natural to require that the matrix $C$ appearing in the output function $y=h(x)$ has full row rank. Yet, not such an assumption {\em explicitly} appears in the statement. We show below that, in fact, the condition is {\em implicitly} encoded in the SDP \eqref{eq:SDP-extended}. To show this, assume by contradiction that $C$ has not full row rank and that \eqref{eq:SDP-extended} is feasible. This implies in particular that there exists $\mathcal{Y}_1$ such that  $\mathcal{Z}_1 \mathcal{Y}_1 +(\mathcal{Z}_1 \mathcal{Y}_1)^\top + 
\alpha I_{n+p}\preceq 0$. We rewrite the matrix on the left-hand side of the inequality as 
\[
\begin{bmatrix}
X_1\\
C \mathcal{Z}_0
\end{bmatrix}
 \mathcal{Y}_1+
 \mathcal{Y}_1^\top 
\begin{bmatrix}
X_1^\top &
\mathcal{Z}_0^\top C^\top 
\end{bmatrix} +\alpha I_{n+p}\preceq 0
\]
where $X_1$ is defined in \eqref{eq:data3}. Let $v\ne 0$ be such that $v^\top C=0$ and multiply the inequality above by $\left[\begin{smallmatrix} 0^\top & v^\top \end{smallmatrix}\right]$ on the left and by $\left[\begin{smallmatrix} 0 \\ v \end{smallmatrix}\right]$ on the right. Then 
\[
v^\top C \mathcal{Z}_0
 \mathcal{Y}_1 \left[\begin{smallmatrix} 0 \\ v \end{smallmatrix}\right] +
\left[\begin{smallmatrix} 0^\top & v^\top \end{smallmatrix}\right]  \mathcal{Y}_1^\top 
\mathcal{Z}_0^\top C^\top v+ \alpha |v|^2\le 0
\]
which is a contradiction. We conclude that $C$ not fulfilling a necessary condition for the feasibility of the regulation problem consistently yields infeasibility of the SDP that is at the basis of the design. 

\smallskip 

{\color{black}It should be kept in mind that the results in this section  can be extended to the case when the dynamics of the system \eqref{sys-w-output} is perturbed by an additive disturbance $w$, which is not necessarily periodic or constant. In this case, the data acquired during the experiment will be affected by $w$ as well. Nonetheless, following the analysis in Section \ref{sec:noisy-data}, if  $w$ satisfies Assumption \ref{ass:RPI}, then one can adjust the SDP in Theorem \ref{thm:contractivity-extended} and obtain a control law 
that makes the closed-loop system \eqref{system-with-disturbance}, \eqref{integral-controller}  contractive. In this way, although the  asymptotic regulation $e(t)\to 0$ as $t\to +\infty$ cannot be guaranteed any longer because of the action of $w(t)$ on the system's response, the closed-loop system will  retain the input-to-state convergence property stated in Corollary \ref{cor2-noisy}.}

\smallskip 

Finally, before considering a numerical example, we observe that, by Corollary \ref{cor:convergent} in the Appendix, a local version of Corollary \ref{cor:integr-control} can be given where $\mathcal{X}$ is a convex subset of  $\mathbb{R}^n$ and 
{\color{black} there exists}
an equilibrium $(\overline x, \overline \xi)$ of the closed-loop system \eqref{extended-system-closed} such that $\overline x\in {\rm int}(\mathcal{X})$. The challenge of applying this result lies in the determination of $\overline x$ due to the unknown disturbance $d$.

\smallskip

\begin{example}
We consider again the single link manipulator with a flexible joint of Example \ref{ex1}, here assuming that the system is corrupted by the constant disturbance $d= \begin{bmatrix} 0.1 & 0.2  & 0.3 & 0.4 \end{bmatrix}^\top$ and $E = I_4$. The regulated output is $y = x_1$ and the constant reference signal is $r = \frac{\pi}{3}$. Note that matrix $C$ satisfies the full row rank condition. The control objective is to regulate the angular position of the link $x_1$ to a desired constant reference $r$. We introduce $Q(x)=  \cos x_1 $. If we set $\mathcal{X} = \mathbb{R}^4$, then \eqref{asspt} is satisfied with $R_Q$ as in \eqref{exmpl1:RQ}.  
We collect $T =10$ samples, i.e. $\mathbb D := \left\{ (x_i,u_i, y_i, \dot x_i, \xi_i )\right\}_{i=0}^{9}$,  by running an experiment with input uniformly distributed in  $[-0.1,0.1]$,
and with an initial state within the same interval. Note that ${\color{black}W=\mathds{1}_{4\times 9}}$. The SDP \eqref{eq:SDP-extended} is feasible and returns the controller ${\color{black}\mathcal{K}}$ and the closed-loop dynamics in \eqref{con.close.integral}. The state evolution of \eqref{close.integral} with initial state uniformly distributed in  $[-1,1]$ is displayed in Figure \ref{plot.integral}, and the closed-loop trajectories converge to an equilibrium point where the tracking error is zero.

\begin{figure}[ht!] \centering
\includegraphics[scale=0.45]{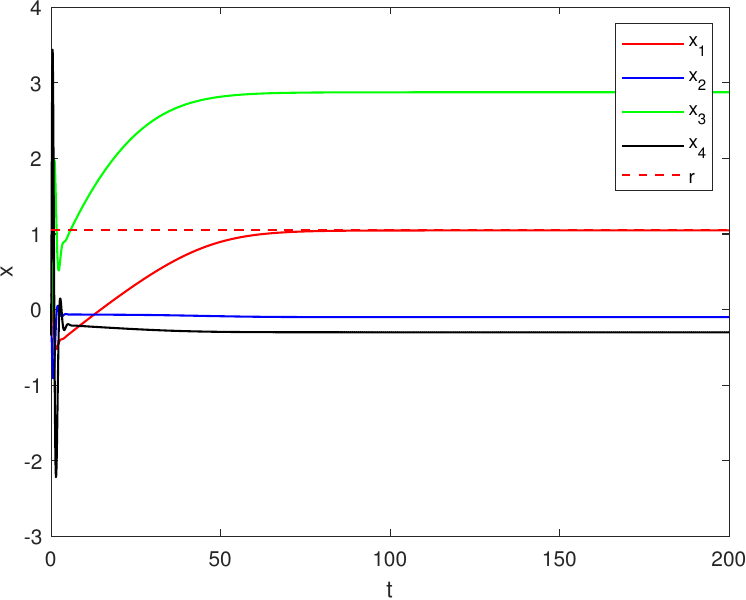}
   \caption{State response of system  \eqref{close.integral} with initial state uniformly distributed in  $[-1,1]$.}\label{plot.integral}
\end{figure}
\smallskip

\begin{figure}[ht!] \centering
\includegraphics[scale=0.45]{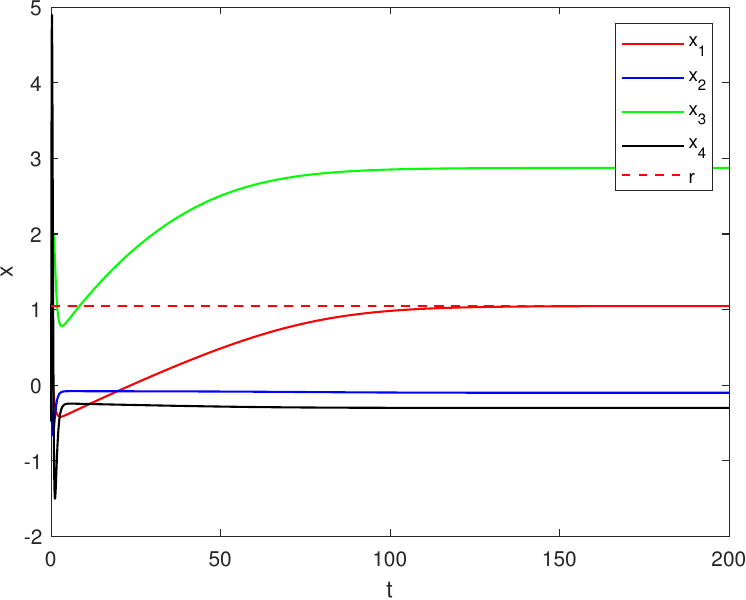}
   \caption{State response of system \eqref{close.integral.more} with initial state uniformly distributed in  $[-1,1]$.}\label{plot.integral.more}
\end{figure}

Next, in order to test the effectiveness of Theorem \ref{thm:contractivity-extended} for other choices of $Q(x)$, let $Q(x)=  \begin{bmatrix} \cos x_1 & \sin x_2 \end{bmatrix}^\top$. Note that 
\[
\frac{\partial Q}{\partial x}(x)^\top \frac{\partial Q}{\partial x}(x) = {\color{black}\text{diag}(\sin(x_1)^2, \cos(x_2)^2, 0, 0)},
\] 
{\color{black} as $x=[x_1\;x_2\;x_3\;x_4\;]^\top$.}
If we set $\mathcal{X} = \mathbb{R}^4$, then \eqref{asspt} is satisfied with 
$
R_Q = R_Q ^\top ={\color{black}\text{diag}(1, 1, 0, 0)}.
$
We collect $T =10$ samples under the same experiment setup. The SDP \eqref{eq:SDP-extended} is feasible and returns the controller ${\color{black}\mathcal{K}}$ and the closed-loop dynamics in \eqref{con.close.integral.more}. The state evolution of \eqref{close.integral.more} with initial state uniformly distributed in  $[-1,1]$ is displayed in Figure \ref{plot.integral.more}, and the closed-loop trajectories converge to an equilibrium point where the tracking error is zero.

\smallskip

\begin{figure*}
\small{\begin{subequations}
\label{con.close.integral}
\begin{align}
 \mathcal{K} &={\color{black} \begin{bmatrix} -5.1682 & -14.5131 &  -7.1368  & -1.6920 &  -0.6164  &  0.0084 \end{bmatrix}}\\ 
{\color{black}\begin{bmatrix} \dot x \\  \dot \xi  \end{bmatrix}} & {\color{black}= }{\color{black}\begin{bmatrix}  0.0000  &  1.0000  &  -0.0000&    -0.0000  &  -0.0000 &  -0.0000\\
   -2.0000  & 0.0000 &   1.0000 &   0.0000  &  0.0000  & -1.9600\\
   -0.0000 &  -0.0000  & -0.0000 &   1.0000 &  -0.0000 &  -0.0000\\
  -33.1210 & -96.7542 & -48.2456 & -11.2800 &  -4.1094  &  0.0559\\
    1.0000 &  -0.0000 &  -0.0000 &  -0.0000 &  -0.0000  &  0.0000\end{bmatrix}} {\color{black}\begin{bmatrix} x \\  \xi \\ \cos x_1  \end{bmatrix} + }{\color{black} \begin{bmatrix} 0.1 \\ 0.2 \\ 0.3 \\ 0.4 \\ 0  \end{bmatrix} - \begin{bmatrix} 0 \\ 0 \\ 0 \\0 \\ \frac{\pi}{3} \end{bmatrix}}\label{close.integral}
\end{align}
\end{subequations}}
\end{figure*}

\begin{figure*}
\small{\begin{subequations}
\label{con.close.integral.more}
\begin{align}
 \mathcal{K} &=\begin{bmatrix} -22.4392 & -96.7758 & -32.6559 &   -4.5347  & -1.6516  &  0.0262  &  0.0000\end{bmatrix}\\ 
{\color{black}\begin{bmatrix} \dot x \\  \dot \xi  \end{bmatrix}} &{\color{black}=} {\color{black}\begin{bmatrix}
   0.0000  &  1.0000  & 0.0000 &  0.0000  & 0.0000  &  -0.0000 &   0.0000\\
   -2.0000  & 0.0000 &   1.0000  &  0.0000  & 0.0000  & -1.9600  & -0.0000\\
   0.0000  &  0.0000  &  0.0000  &  1.0000  & 0.0000 &   -0.0000  & 0.0000\\
 -148.2611 & -645.1721 & -218.3728 & -30.2314 & -11.0107&    0.1745  &  0.0000\\
    1.0000  & -0.0000  & -0.0000  & -0.0000 &   -0.0000  &  -0.0000 &   0.0000\\
\end{bmatrix} \begin{bmatrix} x \\  \xi \\ \cos x_1 \\ \sin x_2 \end{bmatrix} +} {\color{black} \begin{bmatrix} 0.1 \\ 0.2 \\ 0.3 \\ 0.4  \\0\end{bmatrix} - \begin{bmatrix} 0 \\ 0 \\ 0 \\ 0\\ \frac{\pi}{3}  \end{bmatrix}} 
\label{close.integral.more}
\end{align}
\hrulefill
\end{subequations}}
\end{figure*}

\end{example}

%
%

\section{Conclusions}

We have derived data-based semidefinite programs that once solved return nonlinear feedback controllers that guarantee contractivity. For data perturbed by a class of periodic disturbances, these semidefinite programs have been shown to be independent of the magnitude of the disturbances, a remarkable feature. We have also studied the design of data-dependent integral controllers for tracking and disturbance rejection problems. We have chosen to focus on the data-based representation of closed loop systems introduced in \cite{de2019formulas}. {\color{black}We leave to future research the use of} other representations, such as those studied in \cite{bisoffi2022data}, {\color{black}and a comparison between the two approaches.}

%
Although we have considered continuous-time systems, results close to the ones presented here can also be derived for discrete-time systems.  There are manifold options for future work, which could aim at designing controllers for more sophisticated output regulation problems 
\cite{simpson2021lowgain-integral,giaccagli2022sufficient}, working with dictionaries of functions  $Z(x)$ generated via kernels \cite{maddalena2021deterministic,hu2023learning} establishing connections with Gaussian process regression \cite{kawano2023lmi}, and using non-Euclidean norms 
{\color{black}\cite{Proskurnikov2023yakubovich}}, to name a few. 

%

\appendix 
This appendix collects a few basic results on contraction borrowed from the existing literature. 

\subsection{Convergence results for periodic and time-invariant contractive systems}\label{subsec:appA} We consider the system 
\be\label{general.nonlinear.system}
\dot x = f(t,x)
\ee
defined for $(t,x)\in \mathbb{R}\times \mathcal{X}$, where $\mathcal{X}$ is a subset  of $\mathbb{R}^n$. The notation $\phi(t, \tau, \xi)$ indicates the solution of  $\dot x = f(t,x)$, $x(\tau)=\xi$.  It is assumed that $f(t,x)$ is differentiable on $x$ and $f(t,x)$ and $\frac{\partial f}{\partial x}(t,x)$ are continuous in $t,x$. It is also assumed that the set $\mathcal{X}$ is forward invariant, i.e. for any 
$(t_0,x_0)\in \mathbb{R}\times \mathcal{X}$, the solution $\phi(t, t_0, x_0)$ belongs to $\mathcal{X}$ for all $t\in  \mathbb{R}_{\ge t_0}$ that belong to the domain of existence of the solution, and system \eqref{general.nonlinear.system} is forward complete on $\mathcal{X}$,\footnote{Forward completeness is guaranteed if $\mathcal{X}$ is closed and bounded.} i.e. for any $(t_0,x_0)\in \mathbb{R}\times \mathcal{X}$, the solution $\phi(t, t_0, x_0)$ exists for all $t\in \mathbb{R}_{\ge t_0}$.

The following theorem shows the existence of an attractive solution for periodic systems  that are contractive on a subset of the state space, provided that the latter is forward invariant with respect to the system's dynamics and the system's dynamics is forward complete on the set. Claims similar to the one below are given in  \cite[Property 3]{pavlov2005convergent} and \cite[Theorem 3.8]{FB-CTDS} but here we have mostly followed  \cite[Theorem 5]{sontag2010contractive}.

\begin{theorem}\label{lem:sontag}
Assume that
\begin{enumerate}
\item $\mathcal{X}$ is a closed convex subset of $\mathbb{R}^n$;
\item  there exist ${\color{black} P \in \mathbb S^{n \times n}, P \succ 0}, \beta>0$ such that for all 
$(t,x)\in \mathbb{R} \times \mathcal{X}$, 
\[
\displaystyle\frac{\partial f(t,x)^\top}{\partial x} {\color{black}P}^{-1}+{\color{black}P}^{-1}
\displaystyle\frac{\partial f(t,x)}{\partial x}\preceq -\beta {\color{black}P}^{-1}; 
\]
\item for a given $T\in \mathbb{R}_{>0}$, $f$ is $T$-periodic, \emph{i.e.}, $f(t+T, x)=f(t,x)$ for all $(t,x)\in \mathbb{R}\times \mathcal{X}$;
\item $c {\rm e}^{-\beta T/2}<1$, where $c=\left(\frac{\lambda_{\max}({\color{black}P}^{-1})}{\lambda_{\min}({\color{black}P}^{-1})}\right)^{1/2}$. 
\end{enumerate}
Then there exists a unique periodic solution $x_*(t)$ of \eqref{general.nonlinear.system} of period $T$, which is uniformly exponentially stable. 
Furthermore, if 3) is replaced by 
\begin{itemize}
\item [3')] $f(t,x)$ is time invariant, {\emph i.e.}, $f(t,x)=f(x)$ for all $(t,x)\in \mathbb{R}\times \mathcal{X}$, 
\end{itemize}
and 4) is removed, then  there exists a unique equilibrium $x_*$, 
which is uniformly exponentially stable. 
\end{theorem} 

\smallskip

\subsection{Convergent systems}
We recall here the class of convergent systems, which exhibits a solution defined and bounded for all time, to which all the solutions uniformly exponentially converge \cite{pavlov2004convergent}. Compared to the results in the previous subsection, here $\mathcal{X}=\mathbb{R}^n$. 

The system of interest is \eqref{general.nonlinear.system}
defined for $(t,x)\in \mathbb{R}\times \mathbb{R}^n$, with 
$f(t,x)$ differentiable on $x$ and  $f(t,x)$ and $\frac{\partial f}{\partial x}(t,x)$ continuous in $t,x$.

\begin{definition}\label{def:convergent-sys}
The system \eqref{general.nonlinear.system} is convergent if  
\begin{enumerate}
\item all its solutions $x(t)$ are defined for all $t\in {\color{black} \R_{\ge t_0}}$ and all initial conditions $t_0\in \mathbb{R}$, $x(t_0)\in\mathbb{R}^n$;
\item there exists a unique solution $x_*(t)$ defined and bounded for all $t\in {\color{black} \R}$;
\item the solution $x_*(t)$ is globally uniformly exponentially stable.
\end{enumerate}
\end{definition}

\begin{theorem}\label{th:demid-pavlov}
Assume that
\begin{enumerate}
\item $\mathcal{X}=\mathbb{R}^n$;
\item there exist ${\color{black} P \in \mathbb S^{n \times n}, P \succ 0}, \beta>0$ such that for all 
$(t,x)\in \mathbb{R}\times  \mathbb{R}^n$, 
\[
\displaystyle\frac{\partial f(t,x)^\top}{\partial x} {\color{black}P}^{-1}+{\color{black}P}^{-1}
\displaystyle\frac{\partial f(t,x)}{\partial x}\preceq -\beta {\color{black}P}^{-1}; 
\]
\item there exists $\overline x\in \mathcal{X}$ such that, for all $t\in \mathbb{R}$, $|f(t,\overline x)| \le  \bar f <+\infty$.
\end{enumerate}
Then the system is convergent. In addition, if,  for a given $T\in \mathbb{R}_{>0}$, $f$ is $T$-periodic, then $x_*(t)$ is periodic of period $T$. If $f(t,x)$ is time-invariant, then $ x_*(t)=x_*$. 
\end{theorem}

{\emph Proof.} See \cite[Theorem 1]{pavlov2004convergent} and \cite[Property 3]{pavlov2005convergent}. \qedp

\medskip

The proof shows that, denoted by $V(x)$ the function $(x-\overline x)^\top {\color{black}P}^{-1} (x-\overline x)$, it holds that $\dot V(x):=\frac{\partial V}{\partial x} f(t,x)<0$ for all $x\in 
\{x\in \mathbb{R}^n\colon V(x)> \gamma\}$, where $\gamma = (2 \beta^{-1} \bar f  \|{\color{black}P}^{-1/2}\|)^2$. This implies that the set $\mathcal{R}_\gamma =\{x\in \mathbb{R}^n\colon V(x)\le  \gamma\}$ is a closed bounded set of $\mathbb{R}^n$ that is forward invariant  for  $\dot x=f(t,x)$. Hence, it  must  contain  at least one solution $x_*(t)$ defined for all $t\in {\color{black} \R}$ \cite[p.~260]{pavlov2004convergent} (its uniqueness can be shown as in \cite[p.~261]{pavlov2004convergent} and uniform exponential stability is straightforwardly proven).  

The possibility of restricting the result to a convex subset $\mathcal{X}$ of $\mathbb{R}^n$, without explicitly asking for the forward invariance of $\mathcal{X}$ and forward completeness of the system on $\mathcal{X}$, which are the standing assumptions of Theorem \ref{lem:sontag}, depends on whether or not the set $\mathcal{R}_\gamma$ is contained in $\mathcal{X}$, which cannot be in general established, as it depends on $\beta, {\color{black}P}, \bar f $. A special case in which this is true is 
when $\overline x$  is an equilibrium of the system, i.e., $\bar f =0$, and the set $\mathcal{X}$ contains $\overline x$, for in this case 
$ \bar f =0$ implies $\mathcal{R}_\gamma=\{\overline x\}$,  and $x_*(t)=\overline x$. Hence, the following consequence of Theorem \ref{th:demid-pavlov} holds:
\begin{corollary}\label{cor:convergent}
Assume that
\begin{enumerate}
\item $\mathcal{X}$ is a convex subset of  $\mathbb{R}^n$;
\item there exist ${\color{black} P \in \mathbb S^{n \times n}, P \succ 0}, \beta>0$ such that for all 
$(t,x)\in \mathbb{R}\times  \mathcal{X}$, 
\[
\displaystyle\frac{\partial f(t,x)^\top}{\partial x} {\color{black}P}^{-1}+{\color{black}P}^{-1}
\displaystyle\frac{\partial f(t,x)}{\partial x}\preceq -\beta {\color{black}P}^{-1}; 
\]
\item there exists $\overline x\in \text{int}(\mathcal{X})$ such that, for all $t\in \mathbb{R}$, $f(t,\overline x)=0$.
\end{enumerate}
Then, there exists a subset $\mathcal{V}$ of $\mathcal{X}$ containing $\overline x$ such that
\begin{enumerate}
\item all the solutions $x(t)$ of $\dot x=f(t,x)$ are defined for all $t\in {\color{black} \R_{\ge t_0}}$ and all initial conditions $t_0\in \mathbb{R}$, $x(t_0)\in\mathcal{V}$;
\item there exists a unique solution $x_*(t)$ defined and bounded for all $t\in {\color{black} \R}$ and contained in $\mathcal{X}$, and it satisfies $x_*(t)=\overline x$;
\item the solution $x_*(t)=\overline x$ is uniformly exponentially stable.
\end{enumerate}
\end{corollary}

{\it Proof.} Note that  the function $V(x)=(x-\overline x)^\top {\color{black}P}^{-1} (x-\overline x)$ satisfies  $\dot V(x) =\frac{\partial V}{\partial x} f(t,x)<0$ for all $x\in \mathcal{X}$. 
Let $\delta\in \mathbb{R}_{>0}$ and define $\mathcal{V}:=\{x\in \mathbb{R}^n\colon (x-\overline x)^\top {\color{black}P}^{-1} (x-\overline x)\le \delta\}$. Fix $\delta$ such that $\mathcal{V}\subseteq\mathcal{X}$. The set $\mathcal{V}$ is forward invariant with respect to $\dot x=f(t,x)$ and contains a unique  solution $x_*(t)$ defined and bounded for all $t\in {\color{black} \R}$, which satisfies  $x_*(t)=\overline x$. To prove uniqueness, we use the argument of \cite[p.~261]{pavlov2004convergent}. Suppose there exists another solution $x'(t)$ defined and bounded for all $t\in {\color{black} \R}$ and contained in $\mathcal{X}$. Then such a solution satisfies $|x'(t)-\overline x|\le c {\rm e}^{-\beta(t-t_0)/2} |x'(t_0)-\overline x|$ for all $t\in \mathbb{R}_{\ge t_0}$. Letting $t_0\to -\infty$ yields $|x'(t)-\overline x|=0$, which leads to a contradiction by the arbitrariness of $t$. Finally, the solution $x_*(t)=\overline x$ is uniformly exponentially stable as any solution $x(t)$ with initial condition  in $\mathcal{V}$ satisfies 
$|x(t)-\overline x|\le c {\rm e}^{-\beta(t-t_0)/2} |x(t_0)-\overline x|$ for all $t\in \mathbb{R}_{\ge t_0}$. \qedp

\bibliographystyle{IEEEtran}
\bibliography{refs-4}

\vspace{-1cm}



\vspace{-1cm}


\end{document}